%% file: LHC.tex
\documentclass[a4paper]{scrartcl}
\usepackage[T1]{fontenc}
\usepackage[latin1]{inputenc}
\usepackage{graphics}

\makeatletter

\providecommand{\LyX}{L\kern-.1667em\lower.25em\hbox{Y}\kern-.125emX\@}
\newcommand{\noun}[1]{\textsc{#1}}

\input{math.tex}

\def \sw2{\sin^2 \theta_{\rm W}}
\def \cw2{\cos^2 \theta_{\rm W}}

\makeatother

\begin{document}

{\par\raggedleft \texttt{MC-TH-01/13}~\\
 \texttt{MAN/HEP/01/05}~\\
 \texttt{UCL/HEP 2001-}06\\
\texttt{January} \texttt{2002}\par}
\bigskip{}

\bigskip{}
{\par\centering \textbf{\large WW scattering at the LHC}\large \par}
\bigskip{}

{\par\centering J.M. Butterworth\( ^{1} \), \\
 B.E. Cox and J.R. Forshaw\( ^{2} \)\\
\par}
\bigskip{}

{\par\centering {\small \( ^{1} \)Department of Physics \& Astronomy}\\
{\small University College London}\\
{\small Gower St. London WC1E 6BT}\\
{\small England}\small \par}

{\par\centering {\small \( ^{2} \)Department of Physics \& Astronomy }\\
 {\small University of Manchester}\\
 {\small Manchester M13 9PL}\\
{\small England} \par}
\bigskip{}

\vspace{1cm}
{\par\centering \textbf{\small Abstract}\small \par}

{\small A detailed study is presented of elastic \( WW \) scattering in the
scenario that there are no new particles discovered prior to the commissioning
of the LHC. We work within the framework of the electroweak chiral lagrangian
and two different unitarisation protocols are investigated. Signals and backgrounds
are simulated to the final-state-particle level. A new technique for identifying
the hadronically decaying \( W \) is developed, which is more generally applicable
to massive particles which decay to jets where the separation of the jets is
small. The effect of different assumptions about the underlying event is also
studied. We conclude that the channel \( WW\rightarrow jj+l\nu  \) may contain
scalar and/or vector resonances which could be measurable after 100 fb\( ^{-1} \)
of LHC data.}{\small \par}
\newpage

\section{Introduction}

It is quite possible that no new particles will be discovered before the start
of the Large Hadron Collider (LHC). Nevertheless, it is certain that new physics
must reveal itself in or below the TeV region and it is likely that the LHC
will be able to study this new physics in some detail. Precise data collected
at LEP, SLC and the Tevatron interpreted within the Standard Model (or supersymmetric
extensions of the Standard Model) suggest that this new physics should manifest
itself as a higgs boson with mass less than around 200 GeV \cite{lighthiggs}.
However, such a limit is model dependent and it is possible for there to be
no light scalar particle at all \cite{nohiggs,BFS}. 

The scattering of longitudinally polarised vector bosons via the process \( W_{L}W_{L}\rightarrow W_{L}W_{L} \)\footnote{%
We often use the symbol \( W \) to denote both \( W \) and \( Z \) bosons.
} is particularly sensitive to the physics of electroweak symmetry breaking for
it is in this channel that perturbative unitarity is violated at a centre-of-mass
energy of 1.2 TeV. Thus we know that interesting physics must emerge before
then. In the absence of a light higgs, or any other new physics, below some
scale \( \Lambda  \), one can develop a quite general, model independent treatment
of physics well below \( \Lambda  \). This treatment is underpinned by the
electroweak chiral lagrangian (EWChL) \cite{Longhitano}. In this paper we investigate
sensitivity at the LHC to new physics within the EWChL. 

The process \( WW\rightarrow WW \) at high energy hadron colliders has been
studied previously, usually in the context of searches for a heavy Higgs (for
an overview see \cite{TDR,TDR2,ghv,Snowmass}). The \( ZZ \) decay modes constitute
the principal discovery channel for Higgs masses above 160 GeV or so, and the
\( WW \) channels become important around 600 GeV. Within the chiral lagrangian,
it has been usual to focus on leptonic decay modes of the gauge bosons in order
to reduce hadronic backgrounds \cite{bagger,LHCsensitivity}. In this paper
we focus on the more complicated semi-leptonic final state. Cuts developed in
previous studies \cite{TDR2,ghv,zepp,barger,tagjets} are re-examined as a tool
for measuring the cross-section differential in the \( WW \) invariant mass
in the general case (i.e. with no assumption as to the presence or otherwise
of a resonance). A novel technique for identifying the hadronic decays of boosted
massive particles using the longitudinally invariant \( k_{T} \) algorithm
\cite{kt} is introduced, and applied to identification of the hadronically
decaying \( W \). We also examine the sensitivity of the cuts and reconstruction
methods to current simulations of the underlying hadronic activity. 

The paper is set out as follows: The EWChL formalism is introduced in Section
2, and in Section 3 we discuss the unitarisation of the scattering amplitude
for \( W_{L}W_{L}\rightarrow W_{L}W_{L} \). Unitarisation often leads to the
prediction of resonances. We investigate the model dependence of such predictions
and the nature of the resonances (scalar or vector) by looking at two different
unitarisation protocols. In Section 4 we present parton level predictions for
the \( WW \) production cross-section at the LHC for a variety of possible
scenarios. The goal for the LHC will be to distinguish between these different
physics scenarios. To study the potential for this, we have implemented the
general formalism of the EWChL in the \noun{pythia} Monte Carlo program \cite{pythia}.
Sections 5 to 8 cover our analysis of both signal and background. We succeed
in reducing the background to manageable levels using a variety of cuts which
are discussed in detail. Section 9 contains a summary and conclusions.

\section{The Electroweak Chiral Lagrangian}

In the EWChL approach, new physics formally appears in the lagrangian via an
infinite tower of non-renormalisable terms of progressively higher dimension.
However, corrections to observables arising from the new physics can be computed
systematically by truncating the tower at some finite order. This is equivalent
to computing the observable to some fixed order in \( E/\Lambda  \) where \( E \)
is the relevant energy of the experiment.

The breaking of electroweak gauge symmetry already informs us that the scale
of this new physics should be around \( v=246 \) GeV and the degree of symmetry
breaking dictates that our lagrangian should involve three would-be Goldstone
bosons (\( \vec{\pi } \)). Moreover, experiment has told us that after symmetry
breaking there remains, to a good approximation, a residual global \( SU(2) \)
symmetry (often called custodial symmetry) which is responsible for a \( \rho  \)-parameter
of unity (\( \rho =M_{W}^{2}/(M_{Z}^{2}\cw2 ) \)). In chiral perturbation theory
the residual \( SU(2) \) symmetry is the result of the breaking of a global
chiral symmetry, \( SU(2)_{L}\times SU(2)_{R} \). With these constraints, there
is only one dimension-2 term that can be added to the standard electroweak lagrangian
with massless vector bosons. It is\begin{equation}
{\mathcal{L}}^{(2)}=\frac{v^{2}}{4}\langle D_{\mu }UD^{\mu }U^{\dagger }\rangle 
\end{equation}
 where \( \langle \cdots \rangle  \) indicates the \( SU(2) \) trace, and\begin{equation}
U=\exp \left( i\frac{\vec{\pi }\cdot \vec{\tau }}{v}\right) 
\end{equation}
 (\( \vec{\tau } \) are the Pauli matrices). This term contains no physics
that we do not already know. It is responsible for giving the gauge bosons their
mass (this is easiest to see in the unitary gauge where \( U=1 \)). 

At the next order in the chiral expansion, we must include all possible dimension-4
terms. There are only two such terms that will be of relevance to us. They are\begin{equation}
{\mathcal{L}}^{(4)}=a_{4}(\langle D_{\mu }UD^{\nu }U^{\dagger }\rangle )^{2}+a_{5}(\langle D_{\mu }UD^{\mu }U^{\dagger }\rangle )^{2}
\end{equation}
 where \( a_{4} \) and \( a_{5} \) parametrise our ignorance of the new physics
and they are renormalised by one-loop corrections arising from the dimension-2
term. There are a number of additional dimension-4 terms that can arise. However
they generally contribute to anomalous trilinear couplings between vector bosons.
In this paper we focus only on the quartic couplings. In the particular case
of the Standard Model with a heavy higgs boson of mass \( m_{H} \), \( a_{5}=v^{2}/(8m_{H}^{2}) \)
and \( a_{4}=0 \) before renormalisation, whilst for the simplest technicolor
models \( a_{4}=-2a_{5}=N_{TC}/(96\pi ^{2}) \).

To date, other than fixing the scale \( v \) the main constraint on the parameters
of the EWChL come from the precision data on the \( Z^{0} \). Bagger, Falk
\& Swartz have shown that the EWChL can be accommodated without any fine tuning
for \( \Lambda  \) all the way up to 3 TeV (general arguments based on unitarity
indicate that \( \Lambda \lapproxeq 3 \) TeV) \cite{BFS}. They show that the
\( Z^{0} \) data constrain the couplings associated with a dimension-2 custodial
symmetry violating term and a dimension-4 term which contributes to the electroweak
parameter \( S \). There are however no strong constraints on \( a_{4} \)
and \( a_{5} \) and in this paper we assume that they can vary in the range
{[}-0.01,0.01{]} \cite{quartic}.

To one-loop, the EWChL yields the following key amplitude (\( \mu  \) is the
renormalisation scale) \cite{ghv}:\begin{eqnarray}
{\mathcal{A}}(s,t,u) & = & \frac{s}{v^{2}}+\frac{4}{v^{4}}\left[ 2a_{5}(\mu )s^{2}+a_{4}(\mu )(t^{2}+u^{2})+\frac{{1}}{(4\pi )^{2}}\frac{{10s^{2}+13(t^{2}+u^{2})}}{72}\right] \nonumber \\
 &  & -\frac{1}{96\pi ^{2}v^{4}}\left[ t(s+2t)\log (\frac{-t}{\mu ^{2}})+u(s+2u)\log (\frac{-u}{\mu ^{2}})+3s^{2}\log (\frac{-s}{\mu ^{2}})\right] \label{Master} 
\end{eqnarray}
 in terms of which the individual \( W_{L}W_{L}\rightarrow W_{L}W_{L} \) isospin
amplitudes can be written:\begin{eqnarray}
A_{0}(s,t,u) & = & 3A(s,t,u)+A(t,s,u)+A(u,t,s)\\
A_{1}(s,t,u) & = & A(t,s,u)-A(u,t,s)\\
A_{2}(s,t,u) & = & A(t,s,u)+A(u,t,s).\label{isospin} 
\end{eqnarray}
 Equation (\ref{Master}) is derived assuming the Equivalence Theorem wherein
the longitudinal \( W \) bosons are replaced by the Goldstone bosons \cite{ET}.
This approximation is valid for energies sufficiently large compared to the
\( W \) mass.

In addition, (\ref{Master}) is useful only for energies well below \( \Lambda  \),
where the effects of the new physics manifest themselves as small perturbations.
At the LHC, we will be hoping to see much more than small perturbations to existing
physics. For example, we might see new particles associated with the physics
of electroweak symmetry breaking. It would be very useful if we could in some
way extend the domain of validity of the EWChL approach to at least address
the physics that might emerge around the scale \( \Lambda  \). To a degree,
this can be done by invoking some unitarisation protocol which ensures that
(\ref{Master}) develops a high energy behaviour that is consistent with partial
wave unitarity \cite{unitarize}. In the next section, we will consider protocols
that do not spoil the one-loop predictions of the EWChL at lower energies. Such
an approach has met with some success in extending studies of chiral perturbation
theory in QCD \cite{QCD}. We will focus on two unitarisation protocols: the
Pad\'{e} protocol and the \( N/D \) protocol.

\section{Unitarisation}

The amplitude in the weak isospin basis, \( A_{I} \), can be projected onto
partial waves, \( t_{IJ} \), with definite angular momentum \( J \) and weak
isospin \( I \):\begin{equation}
\label{tIJ}
t_{IJ}=\f {1}{64\pi }\int _{-1}^{1}d(\cos \theta )\, P_{J}(\cos \theta )\, A_{I}(s,t,u)
\end{equation}

where \( \theta  \) is the centre-of-mass scattering angle. The \( WW \) scattering
system can have \( I=0,1,2 \) and Bose symmetry further implies that only even
\( J \) are allowed for \( I=0 \) and 2, while only odd \( J \) are allowed
for \( I=1 \)\textit{.} Subsequently we consider the three amplitudes \( t_{00},t_{11},t_{20} \).
The higher partial waves are strictly of order \( s^{2}/v^{4} \) but they are
numerically small and we neglect them. 

Writing \( t_{IJ}=t_{IJ}^{(2)}+t_{IJ}^{(4)}+\cdots  \) , the first two terms
of the expansion are given by \cite{LHCsensitivity}:\begin{equation}
\label{t002}
t_{00}^{(2)}=\f {s}{16\pi v^{2}}
\end{equation}

\begin{equation}
\label{t004}
t_{00}^{(4)}=\f {s^{2}}{64\pi v^{4}}\left[ \f {16(11a_{5}(\mu )+7a_{4}(\mu ))}{3}+\f {1}{16\pi ^{2}}\left( \f {101-50\log (s/\mu ^{2})}{9}+4i\pi \right) \right] 
\end{equation}

\begin{equation}
\label{t112}
t_{11}^{(2)}=\f {s}{96\pi v^{2}}
\end{equation}

\begin{equation}
\label{t114}
t_{11}^{(4)}=\f {s^{2}}{96\pi v^{4}}\left[ 4(a_{4}(\mu )-2a_{5}(\mu ))+\f {1}{16\pi ^{2}}\left( \f {1}{9}+\f {i\pi }{6}\right) \right] 
\end{equation}

\begin{equation}
\label{t202}
t_{20}^{(2)}=-\f {s}{32\pi v^{2}}
\end{equation}

\begin{equation}
\label{t204}
t_{20}^{(4)}=\f {s^{2}}{64\pi v^{4}}\left[ \f {32(a_{5}(\mu )+2a_{4}(\mu ))}{3}+\f {1}{16\pi ^{2}}\left( \f {91}{18}-\f {20\log (s/\mu ^{2})}{9}+i\pi \right) \right] .
\end{equation}

Using\begin{equation}
\label{partial}
A_{I}(s,t,u)=32\pi \, \sum ^{\infty }_{J=0}(2J+1)t_{IJ}\, P_{J}(\cos \theta )
\end{equation}
 we have (neglecting higher partial waves)\begin{eqnarray}
A_{0}(s,t,u) & = & 32\pi \, t_{00}\nonumber \\
A_{1}(s,t,u) & = & 32\pi \, 3t_{11}\cos \theta \nonumber \\
A_{2}(s,t,u) & = & 32\pi \, t_{20}.
\end{eqnarray}
In terms of these amplitudes we can write\begin{eqnarray}
A(W^{+}W^{-}\to W^{+}W^{-}) & = & \frac{1}{3}A_{0}+\frac{1}{2}A_{1}+\frac{1}{6}A_{2}\nonumber \\
A(W^{+}W^{-}\to ZZ) & = & \frac{1}{3}A_{0}-\frac{1}{3}A_{2}\label{Wamps} \\
A(ZZ\to ZZ) & = & \frac{1}{3}A_{0}+\frac{2}{3}A_{2}\nonumber \\
A(WZ\to WZ) & = & \frac{1}{2}A_{1}+\frac{1}{2}A_{2}\nonumber \\
A(W^{\pm }W^{\pm }\to W^{\pm }W^{\pm }) & = & A_{2}.\nonumber 
\end{eqnarray}
 The differential \( WW \) cross-section is\begin{equation}
\label{WWxsecn}
\frac{d\sigma }{d\cos \theta }=\frac{|A(s,t)|^{2}}{32\pi \, M_{WW}^{2}}.
\end{equation}

To obtain the cross-section for \( pp\to WWjj+X \) we need to fold in the parton
density functions, \( f_{i}(x,Q^{2}) \), and the \( WW \) luminosity: \begin{equation}
\label{ppxsecn}
\frac{d\sigma }{dM_{WW}^{2}}=\sum _{i,j}\int ^{1}_{M_{WW}^{2}/s}\int _{M_{WW}^{2}/(x_{1}s)}^{1}\frac{dx_{1}\, dx_{2}}{x_{1}x_{2}s_{pp}}f_{i}(x_{1},M_{W}^{2})\, f_{j}(x_{2},M_{W}^{2})\frac{dL_{WW}}{d\tau }\int ^{1}_{-1}\frac{d\sigma }{d\cos \theta }d\cos \theta 
\end{equation}

where \( \surd s_{pp} \) is the centre-of-mass energy which we take to be 14
TeV, as appropriate for the LHC, \[
\frac{dL_{WW}}{d\tau }\approx \left( \frac{\alpha }{4\pi \sin ^{2}\theta _{W}}\right) ^{2}\frac{1}{\tau }\left[ (1+\tau )\ln (1/\tau )-2(1-\tau )\right] \]

for incoming \( W^{\pm } \) bosons \cite{EWA} and \( \tau =M_{WW}^{2}/(x_{1}x_{2}s_{pp}) \).

\paragraph*{The Pad\'{e} protocol}

Otherwise known as the Inverse Amplitude Method, this is a simple unitarisation
procedure, and is widely employed \cite{LHCsensitivity,Pade,mms}. Elastic unitarity
demands that for \( s>0 \) the imaginary part of the amplitude is equal to
the modulus squared of the amplitude, which implies\begin{equation}
t_{IJ}^{-1}=\re (t_{IJ}^{-1})-i.
\end{equation}

To the accuracy in which we work, we can write\begin{equation}
\label{Pade}
t_{IJ}=\frac{t_{IJ}^{(2)}}{\left( 1-\frac{t_{IJ}^{(4)}(s)}{t_{IJ}^{(2)}(s)}\right) }
\end{equation}
 which has the virtue that it satisfies the elastic unitarity condition identically.
We stress that this method of unitarisation leads to an amplitude that is equivalent
to the one-loop EWChL calculation modulo higher-orders in \( s/v^{2} \). 

Having unitarised the amplitude it is natural to ask what the consequences are.
Typically, the partial waves develop resonances which serve to implement the
demands of unitarity; this is the role played by the Higgs boson in the Standard
Model. The position and nature of the resonances depends critically upon the
unitarisation protocol and we investigate an alternative protocol in the following
subsection. At high enough energy, the partial waves effectively lose all memory
of the underlying chiral perturbation theory and their nature is driven solely
by the choice of unitarisation protocol. We therefore rely on our unitarisation
protocol to provide us with some feeling for the pattern of lowest lying resonances
which may be observed at future colliders. 

Resonances are found whenever the corresponding phase shift passes through \( \pi /2 \),
i.e. when \[
\cot \delta _{IJ}=\re (t_{IJ}^{-1})=0.\]

\begin{figure}
{\par\centering \resizebox*{0.6\textwidth}{!}{\includegraphics{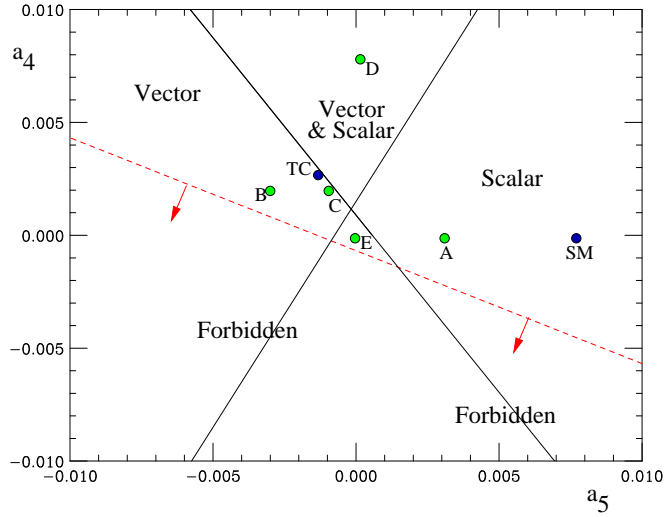}} \par}

\caption{Map of the parameter space as determined by the Pad\'e protocol \cite{LHCsensitivity}.
The small triangle in the centre is the region of no resonances and the region
below the dotted line is forbidden. Also shown are the points corresponding
to the various scenarios considered in the text.\label{PadeMap}}
\end{figure}
In the Pad\'e approach we can solve this equation to obtain the corresponding
masses and widths \cite{LHCsensitivity}. For scalar resonances:\begin{equation}
\label{scalar0}
m_{S}^{2}=\frac{{4v^{2}}}{\frac{{16}}{3}(11a_{5}(\mu )+7a_{4}(\mu ))+\frac{{1}}{16\pi ^{2}}\left( \frac{{101-50\log (m_{S}^{2}/\mu ^{2})}}{9}\right) }
\end{equation}
and\[
\Gamma _{S}=\frac{{m_{S}^{3}}}{16\pi v^{2}}.\]
For vector resonances:\begin{equation}
\label{vector1}
m_{V}^{2}=\frac{{v^{2}}}{4(a_{4}(\mu )-2a_{5}(\mu ))+\frac{{1}}{16\pi ^{2}}\frac{{1}}{9}}
\end{equation}

and\[
\Gamma _{V}=\frac{{m_{V}^{3}}}{96\pi v^{2}}.\]
There are no resonances in the isotensor channel, i.e. from \( t_{20} \). There
is however a region of parameter space where the phase shift passes through
\( -\pi /2 \). This would violate causality and so we are forced to forbid
such regions of parameter space. It occurs when \begin{equation}
\label{forbid}
\frac{{32}}{3}(a_{5}(\mu )+2a_{4}(\mu ))+\frac{1}{16\pi ^{2}}\left( \frac{273}{54}-\frac{20}{9}\log \frac{s}{\mu ^{2}}\right) <0.
\end{equation}

A map of the \( a_{4}-a_{5} \) parameter space showing the corresponding resonance
structure is presented in Figure \ref{PadeMap}. We fix \( \mu =1 \) TeV and,
using equations (\ref{scalar0}) and (\ref{vector1}), we define the regions
to contain a resonance of the specified type with mass below 2 TeV. The blue
points labelled TC and SM correspond to the naive \( N_{TC}=3 \) technicolor
(TC) and 1 TeV Standard Model Higgs (SM) models.

\paragraph*{The \protect\( N/D\protect \) protocol }

This provides our alternative to the Pad\'{e} protocol. This method ensures
that the amplitude has improved analytic properties in addition to satisfying
partial wave unitarity and matching the one-loop EWChL calculation. The right-hand
cut is placed wholly into the denominator function, \( D \), while the left-hand
cut is encapsulated in the numerator function, \( N \), i.e. analyticity and
unitarity demand the following relations \cite{mms,oller,hi}:\begin{eqnarray}
\im (t_{IJ}(s)^{-1})=-1\qquad s>0 &  & \label{NoverD1} \\
\im \, D(s)=0\qquad s<0 &  & \label{NoverD2} \\
\im \, N(s)=D\, \im \, t_{IJ}(s)\qquad s<0 &  & \label{NoverD3} \\
\im \, N(s)=0\qquad s>0 & \label{NoverD4} 
\end{eqnarray}
 where \begin{equation}
t_{IJ}(s)=\frac{N(s)}{D(s)}.
\end{equation}
Following Oller, we define the following function to contain the right-hand
cut at \( s=M^{2} \) \cite{oller}: \begin{equation}
g(s)=\f {1}{\pi }\log \left( -\f {s}{M^{2}}\right) 
\end{equation}
 where \( M \) is an unknown parameter. The \( N/D \) unitarised partial wave
amplitude is then written \begin{equation}
t_{IJ}(s)=\frac{X_{IJ}(s)}{1+g(s)X_{IJ}(s)}
\end{equation}

where \begin{equation}
X_{IJ}(s)=t^{(2)}_{IJ}(s)+t^{(4)}_{IJ}(s)+g(s)(t^{(2)}_{IJ}(s))^{2}.
\end{equation}

The amplitude thus defined has been constructed so as to satisfy (\ref{NoverD1})
and (\ref{NoverD4}) identically whilst (\ref{NoverD2}) and (\ref{NoverD3})
are satisfied to one-loop in chiral perturbation theory. Note that the contribution
to \( \im \, D(s) \) for \( s<0 \) is beyond the one-loop approximation.
\begin{figure}
{\par\centering \resizebox*{0.8\columnwidth}{!}{\includegraphics{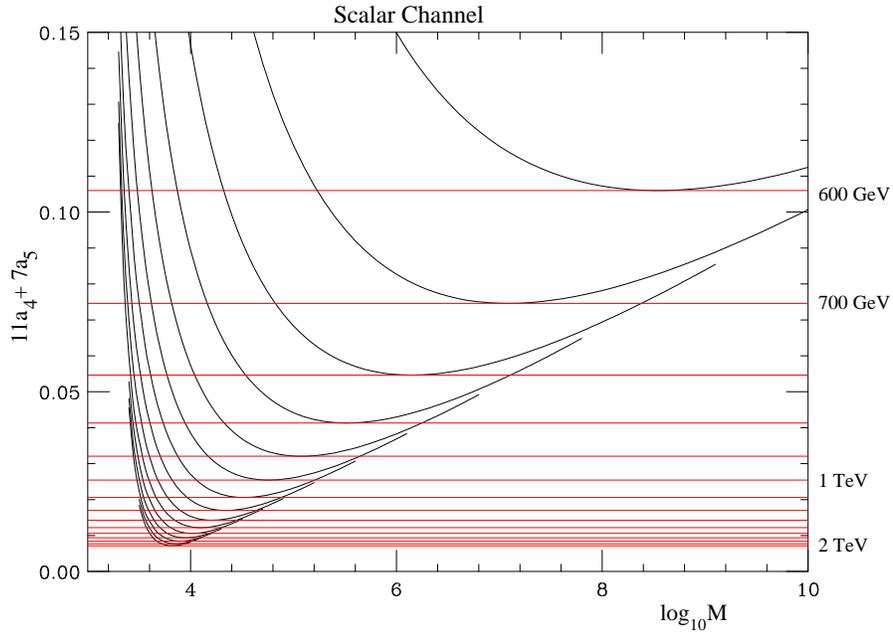}} \par}

\caption{Contours of constant scalar resonance mass in steps of 100 GeV. The horizontal
lines are obtained using the Pad\'e protocol and the curved lines are obtained
using the N/D method which depends upon the parameter \protect\( M\protect \).\label{NDscalar}}
\end{figure}

\begin{figure}
{\par\centering \resizebox*{0.8\columnwidth}{!}{\includegraphics{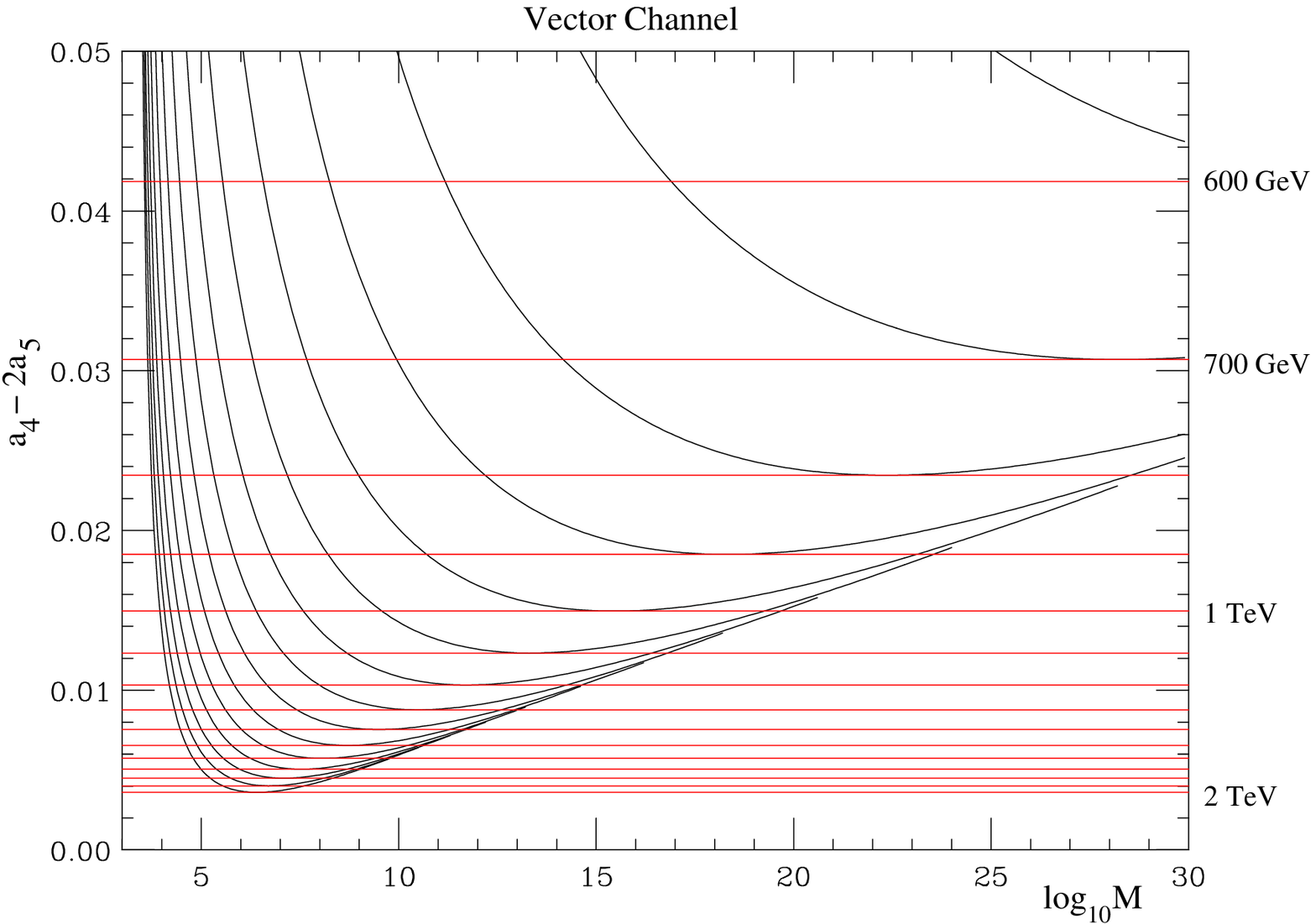}} \par}

\caption{Contours of constant vector resonance mass in steps of 100 GeV. The horizontal
lines are obtained using the Pad\'e protocol and the curved lines are obtained
using the N/D method which depends upon the parameter \protect\( M\protect \).\label{NDvector}}
\end{figure}

\begin{figure}
{\par\centering \resizebox*{0.8\columnwidth}{!}{\includegraphics{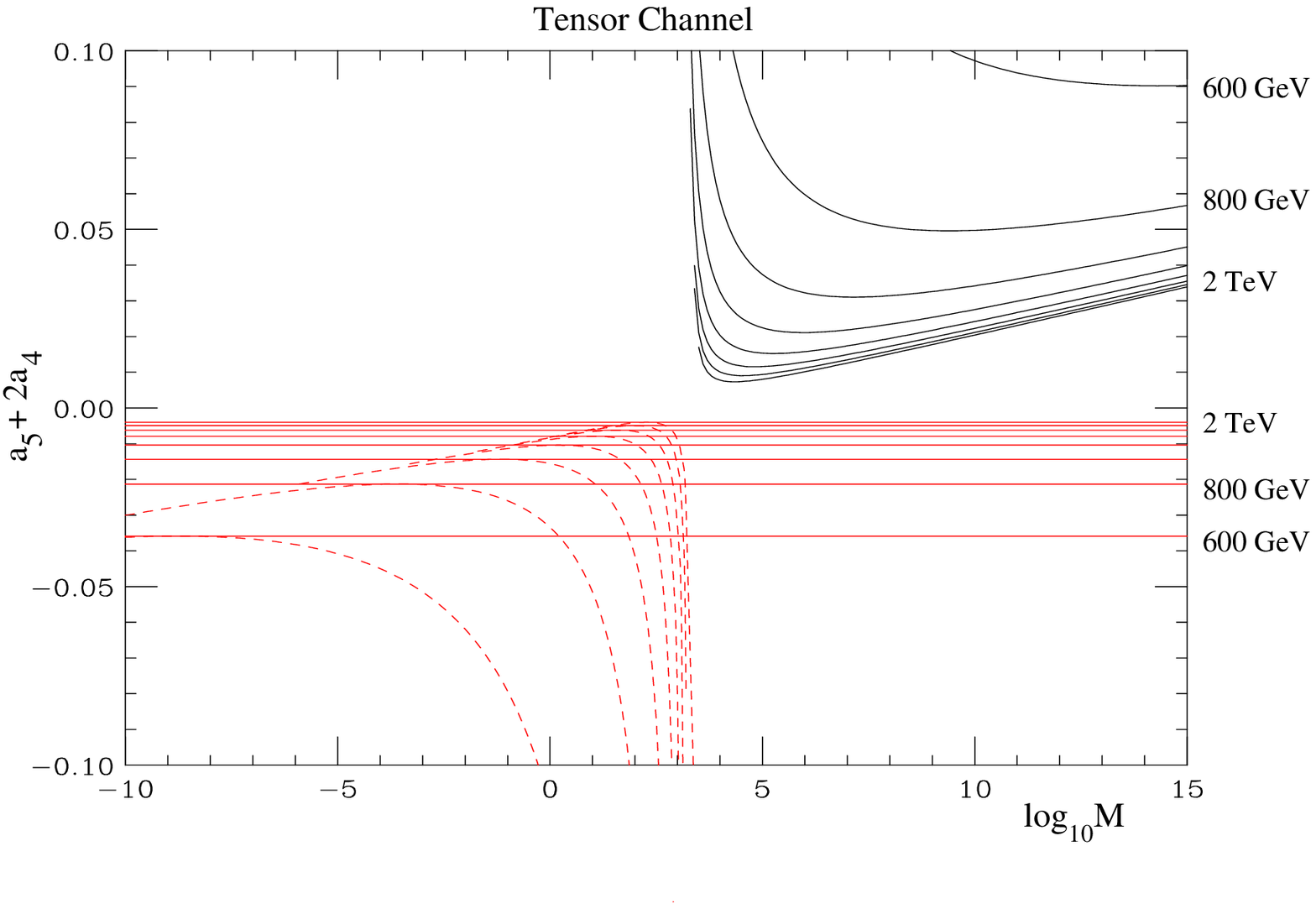}} \par}

\caption{Contours of constant isotensor resonance mass in steps of 100 GeV. The curved
lines are obtained using the N/D method which depends upon the parameter \protect\( M\protect \).
The horizontal lines and dotted curves populate the unphysical region of parameter
space, see the discussion in the text.\label{NDtensor}}
\end{figure}
In Figures \ref{NDscalar} to \ref{NDtensor} we show curves of constant resonance
mass, varying from 600 GeV to 2 TeV in steps of 100 GeV, as a function of the
appropriate combination of \( a_{4}(1 \)~TeV), \( a_{5}(1 \)~TeV), and \( M \).
The horizontal lines obtained using the Pad\'e protocol are tangent to the corresponding
N/D contours. Over large regions of parameter space, the two protocols yield
similar results. However, the N/D method predicts a larger region without resonances,
indeed for \( M \) below around 1 TeV there are no resonances at all. Referring
back to Figure \ref{PadeMap}, we see that as \( M \) increases the lines which
define the scalar and vector regions move slowly outwards. From Figure \ref{NDtensor}
we see that for \( M \) above \( \sim 1 \) TeV the region excluded in the
Pad\'e protocol is not excluded in the N/D protocol leading instead to a region
without any resonances. For \( M \) below \( \sim 1 \) TeV, there is a region
excluded by N/D unitarisation (for the same reason as in the earlier Pad\'e
case)\footnote{%
The scalar and vector sectors have exclusion regions similar to the tensor sector.
}. The line delineating the forbidden region in Figure \ref{PadeMap} thus moves
slowly downwards as \( M \) decreases. Note that we do not know the natural
value for \( M \), e.g. it can be much smaller than 1 TeV. Finally, we note
from Figure \ref{NDtensor} that the N/D method does allow for the existence
of doubly charged resonances.

\section{Parton Level Predictions for the LHC}

\begin{figure}
{\par\centering \resizebox*{0.8\columnwidth}{!}{\rotatebox{90}{\includegraphics{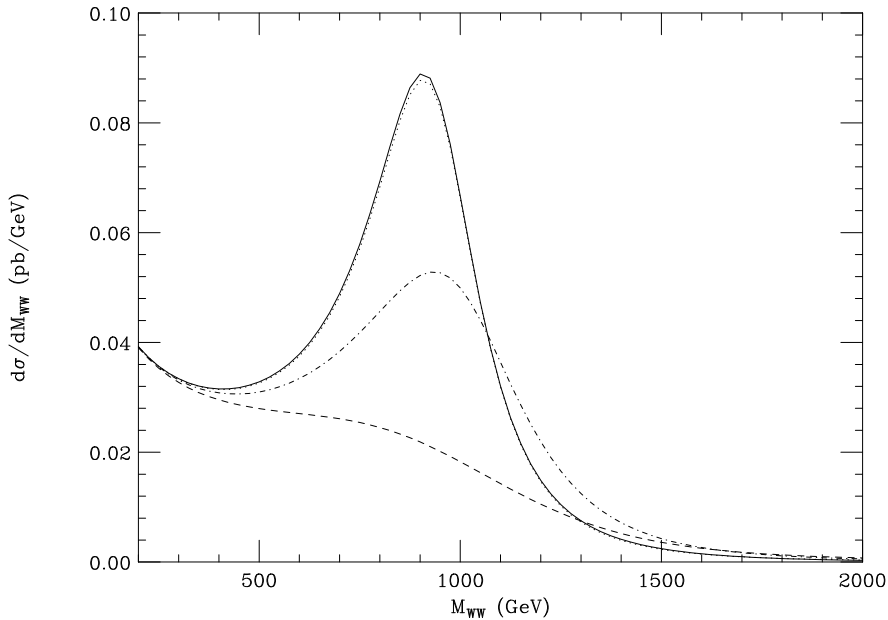}}} \par}

\caption{Parton level cross-section for Scenario A. We compare the Pad\'e result (solid
line) with the N/D results for \protect\( M=10^{3}\protect \) GeV(dashed line),
\protect\( M=10^{4}\protect \) GeV (dashed-dotted line) and \protect\( M=10^{5}\protect \)
GeV (dotted line).\label{PartonA}}
\end{figure}

\begin{figure}
{\par\centering \resizebox*{0.8\columnwidth}{!}{\rotatebox{90}{\includegraphics{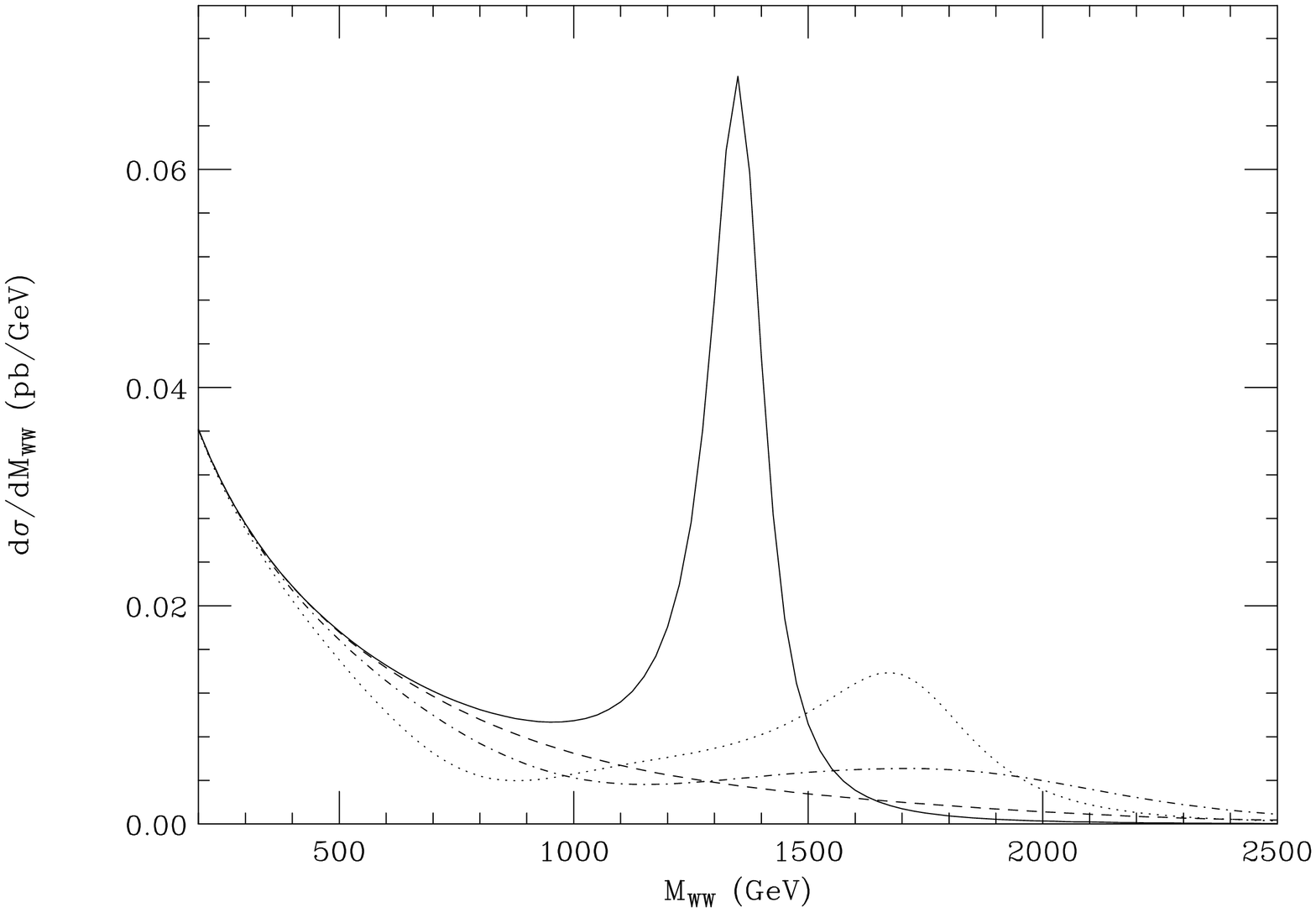}}} \par}

\caption{Parton level cross-section for Scenario B. We compare the Pad\'e result with
the N/D results as in Figure \ref{PartonA}.\label{PartonB}}
\end{figure}

\begin{figure}
{\par\centering \resizebox*{0.8\columnwidth}{!}{\rotatebox{90}{\includegraphics{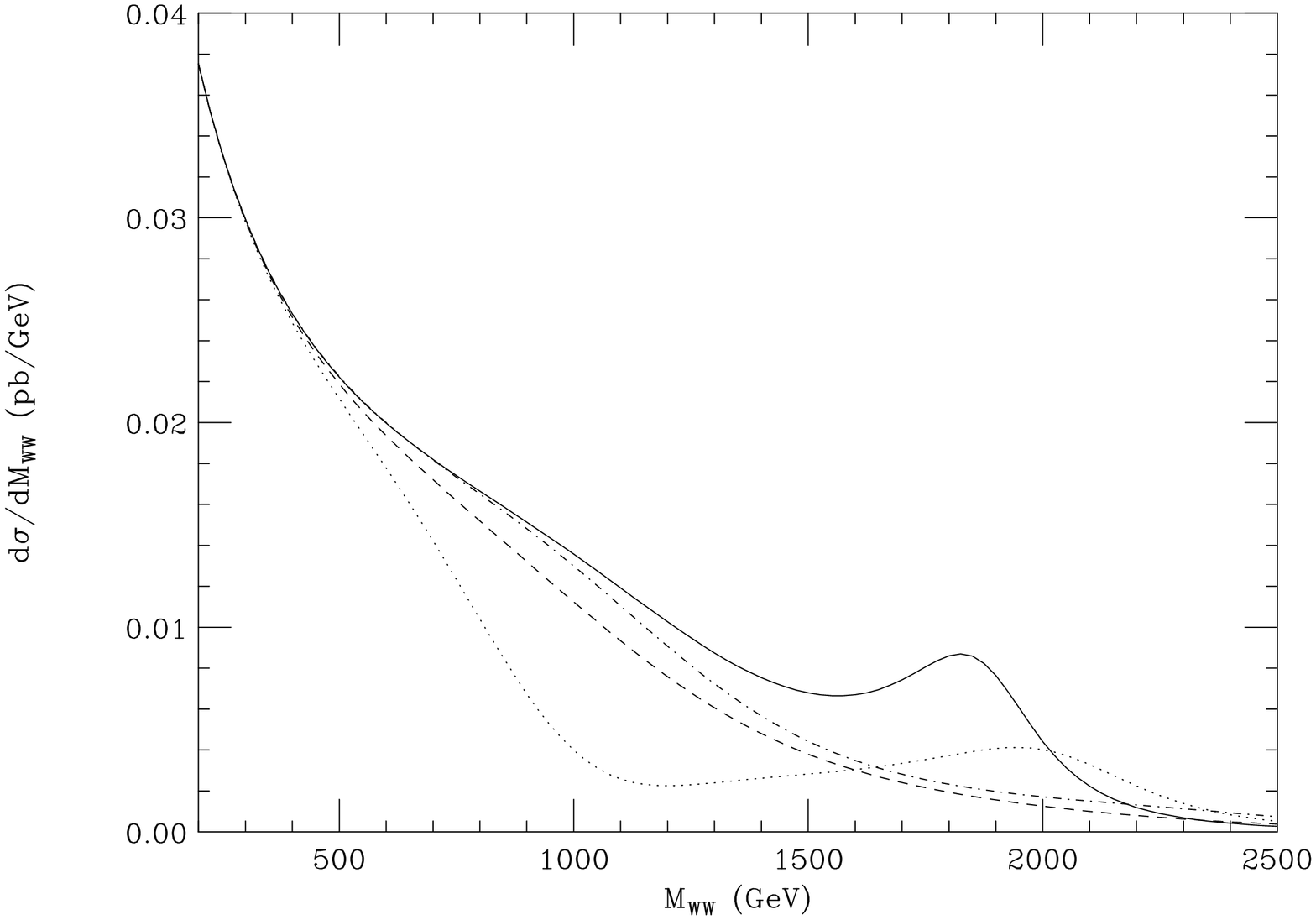}}} \par}

\caption{Parton level cross-section for Scenario C. We compare the Pad\'e result with
the N/D results as in Figure \ref{PartonA}.\label{PartonC}}
\end{figure}

\begin{figure}
{\par\centering \resizebox*{0.8\columnwidth}{!}{\rotatebox{90}{\includegraphics{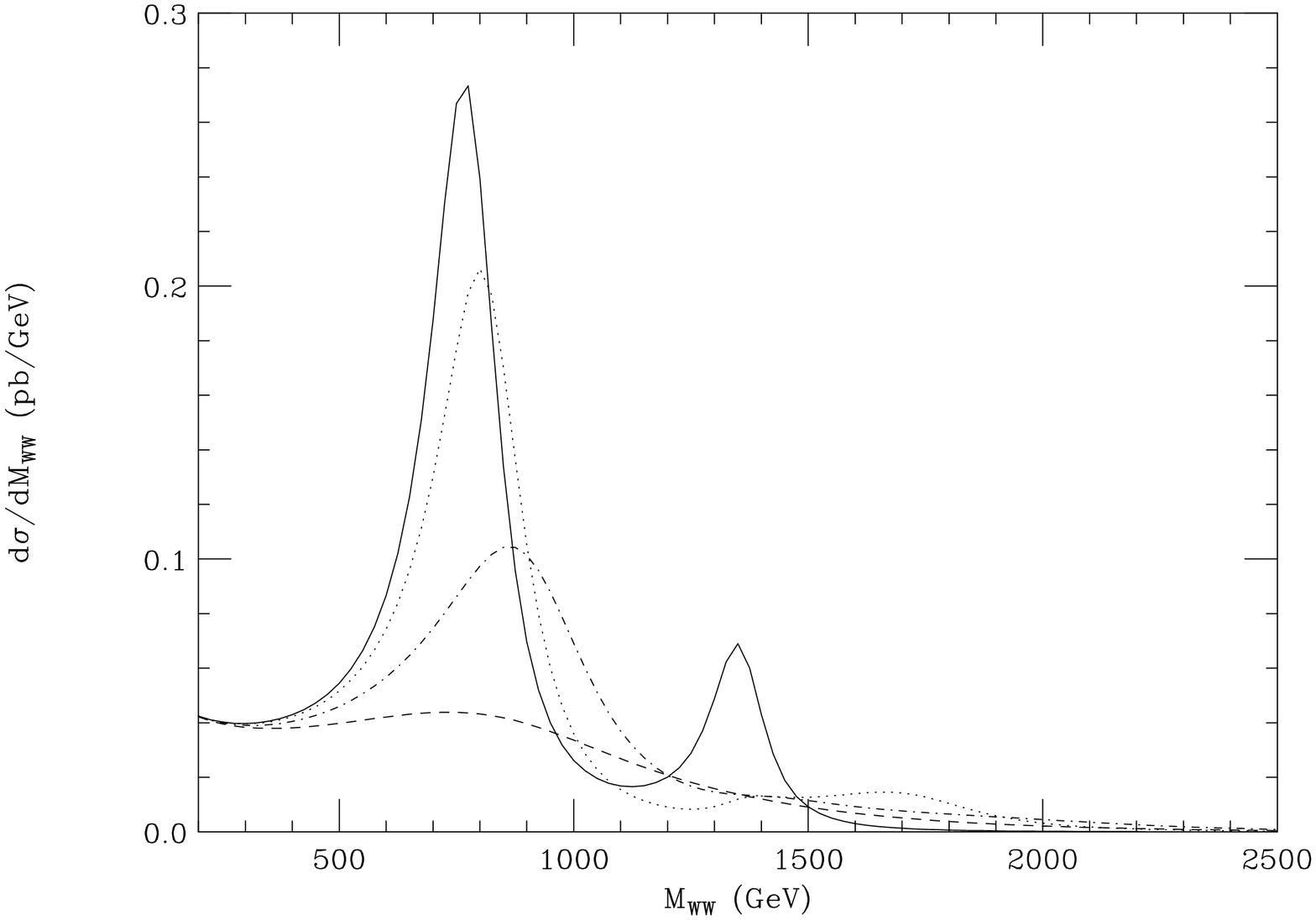}}} \par}

\caption{Parton level cross-section for Scenario D. We compare the Pad\'e result with
the N/D results as in Figure \ref{PartonA}.\label{PartonD}}
\end{figure}

\begin{figure}
{\par\centering \resizebox*{0.8\columnwidth}{!}{\rotatebox{90}{\includegraphics{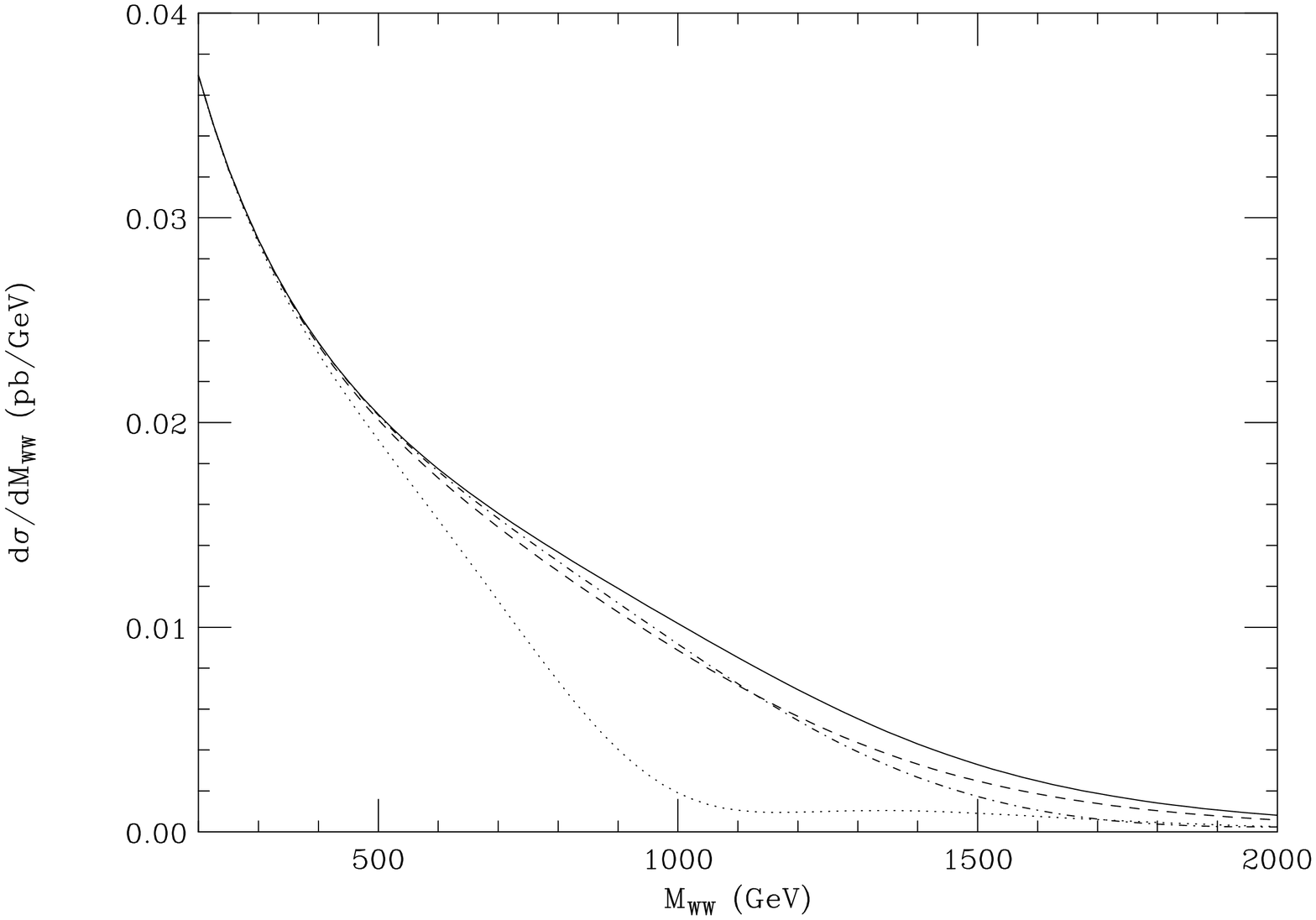}}} \par}

\caption{Parton level cross-section for Scenario E. We compare the Pad\'e result with
the N/D results as in Figure \ref{PartonA}.\label{PartonE}}
\end{figure}

\begin{table}
{\centering \begin{tabular}{|c|c|c|}
\hline 
Scenario&
\( a_{4} \)(1 TeV)&
\( a_{5} \)(1 TeV)\\
\hline 
\hline 
A&
0.0&
0.003\\
\hline 
B&
0.002&
-0.003\\
\hline 
C&
0.002&
-0.001\\
\hline 
D&
0.008&
0\\
\hline 
E&
0&
0\\
\hline 
\end{tabular}\par}

\caption{Parameters for the five scenarios which we consider.\label{PramTable}}
\end{table}
In this section the parton level predictions for the process \( pp\to W^{+}W^{-}jj+X \)
at 14 TeV centre-of-mass energy are presented for the 5 different choices of
\( a_{4} \) and \( a_{5} \) shown in Table \ref{PramTable}. In the Pad\'e
approach these choices produce a 1 TeV scalar (scenario A), a 1 TeV vector (scenario
B), a 1.9 TeV vector (scenario C), a 800 GeV scalar and a 1.4 TeV vector (scenario
D), and a scenario with no resonances (scenario E). The green points labelled
A-E on Figure \ref{PadeMap} correspond to the 5 scenarios we consider. Throughout
this paper the CTEQ4L \cite{cteq} parton density functions as implemented in
PDFLIB \cite{PDFLIB} are used, evaluated at the \( WW \) centre-of-mass energy
\( (M_{WW}) \). The renormalisation scale is fixed to 1 TeV. The differential
cross-section \( d\sigma /dM_{WW} \) for each of scenarios A-E are shown in
Figures \ref{PartonA}-\ref{PartonE}. We compare the Pad\'e protocol with results
using the N/D protocol for three different values of the mass parameter \( M \).

Note that, for values of \( M \) below around 10 TeV there are no resonances
at all in the N/D scenario. This is in accord with expectations based on Figures
\ref{NDscalar} to \ref{NDtensor}. Also, if \( M \) becomes too large then
it leads to unusual behaviour of the amplitudes due to the dominance of the
\( g(s) \) term which suppresses the amplitudes away from the region of resonances
and can produce zeros in the individual partial wave amplitudes. The tail in
the dotted line shown in Figure \ref{PartonE} is a consequence of such behaviour.
Just discernable in Figure \ref{PartonD} is an isospin 2 scalar resonance just
below 1.5 TeV in the N/D dotted curve.

\section{Monte Carlo Simulations}

We have modified the \textsc{Pythia} Monte Carlo generator \cite{pythia} to
include the EWChL approach using both Pad\'e and N/D protocols. Signal samples
containing the \( W^{\pm }W^{\pm } \) final state (including all charge combinations)
are generated using \textsc{Pythia} 6.146 with the Pad\'e unitarisation scheme\footnote{%
The code is available from the authors on request.
}. As a cross check, a sample with a 1 TeV Higgs was also generated using \textsc{Herwig}
6.1 \cite{herwig}.
\begin{figure}
{\par\centering \resizebox*{0.85\columnwidth}{!}{\includegraphics{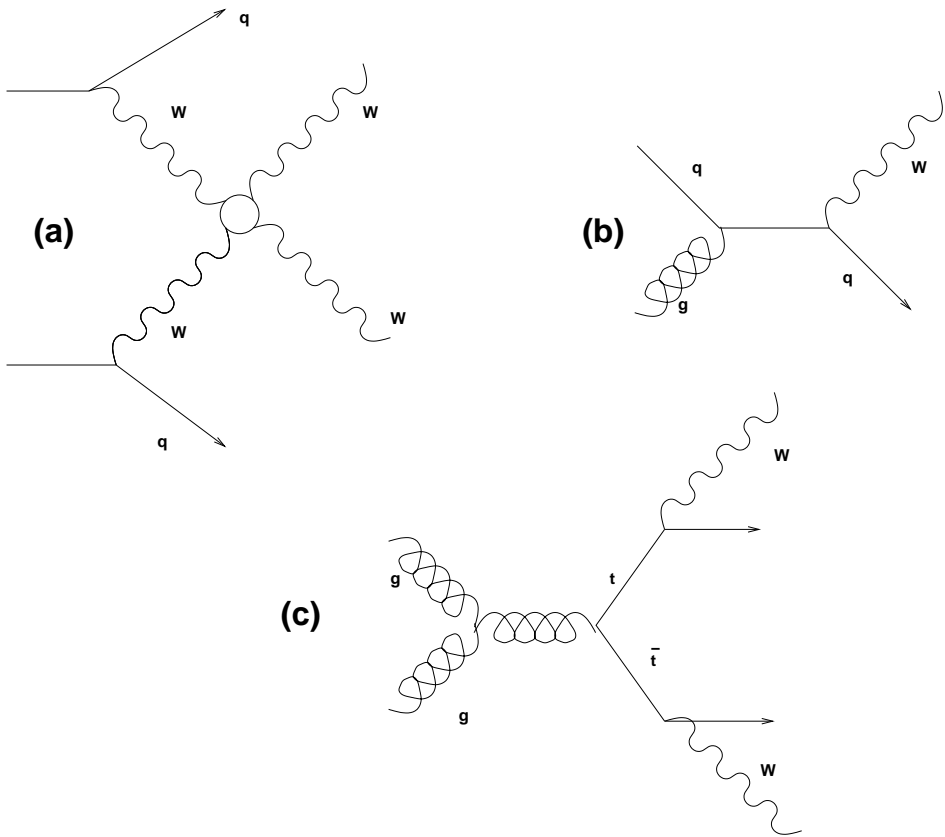}} \par}

\caption{Typical diagrams for signal and background processes: (a) signal; (b) \protect\( W+\protect \)jets;
(c) \protect\( t{}\bar{t}\protect \)\label{Feynman}}
\end{figure}

The dominant backgrounds are QCD \( t\bar{t} \) production and radiative \( W+ \)jets,
as illustrated in Figure \ref{Feynman}. These processes are implemented in
the \textsc{Pythia} 6.146 and \textsc{Herwig 6.1} generators. To improve generation
efficiency the minimum \( p_{T} \) of the hard scatter is set to 250 GeV for
the \( W+ \) jets sample and to 300 GeV for the \( t\bar{t} \) sample \cite{TDR2}.
In addition to the hard subprocesses, the effects of the {}``underlying event{}''
are simulated in both signal and background. Our default model in \textsc{Pythia}
\cite{zijl} is obtained by setting a fixed minimum \( p_{T} \) cut off of
\( p_{T}^{min}=3 \) GeV for secondary scatters. The default energy dependence
of this cut-off has been explicitly turned off. No pile-up from multiple \( pp \)
interactions is included. Other models, in both \textsc{Herwig} and \textsc{Pythia},
are discussed in Section \ref{UE}, along with their effects. The leading order
cross-sections are used to obtain rates and there is therefore a rather large
degree of uncertainty, particularly in \( t\bar{t} \) production, which is
a pure QCD, dominantly gluon induced, process. NLO calculations \cite{ttNLO}
suggest K-factors of order two are appropriate; the final word would come from
measurements at the LHC itself.

\section{Extracting the Signal\label{extract} }

To identify semileptonic decays, we select first on the leptonically decaying
\( W \) (electron/muon and missing transverse energy), then on the hadronically
decaying \( W \) (jet invariant mass, rapidity and transverse energy) and finally
on the event environment (tagging jets at high rapidities, vetoing on central
minijet activity). In all cases we have used only particles within a rapidity
region of \( |\eta |<4.5 \) to approximate the acceptance of a general purpose
detector at the LHC. For clarity, we show just one signal sample as an example.
The 1 TeV scalar resonance (scenario A) is chosen, since this has the lowest
average \( M_{WW} \) and therefore has a shape closest to that of the backgrounds.
The other scenarios, while in general very like this sample, have a harder spectrum
in the transverse momentum variables. The analysis follows the 1 TeV Higgs study
of \cite{TDR2} quite closely for many cuts. However, we differ in the identification
of hadronically decaying \( W \) bosons via the subjet method, in the top quark
veto, in the cut on the transverse momentum of the hard system, and in details
of other cuts; all of which are described below.

\subsection{Leptonic Variables}

\begin{figure}
{\par\centering \resizebox*{1\textwidth}{!}{\includegraphics{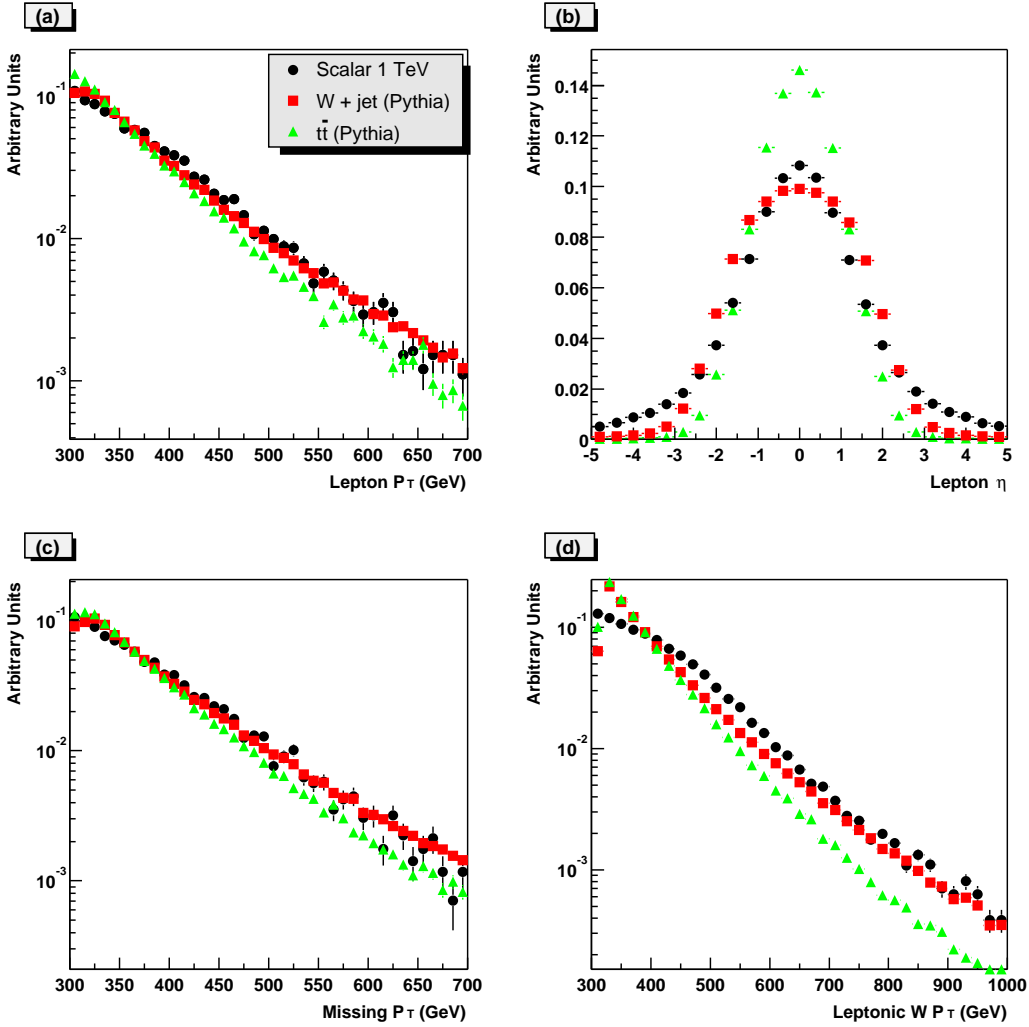}} \par}

\caption{\label{leptons}Leptonic variables for signals and background (a) transverse
momentum of the highest \protect\( p_{T}\protect \) charged lepton (\protect\( e\protect \)
or \protect\( \mu \protect \)), (b) pseudorapidity of the same lepton, (c)
missing transverse momentum and (d) the \protect\( p_{T}\protect \) of the
\protect\( W\protect \) candidate constructed from the lepton and the assumed
neutrino. The area under the histograms is set to one to allow comparison of
the shapes. A trigger cut at 80 GeV in the \protect\( p_{T}\protect \) of the
highest \protect\( p_{T}\protect \) jet and at 40 GeV in the highest \protect\( p_{T}\protect \)
charge lepton is applied before making the plots, as well as a realistic rapidity
acceptance.}
\end{figure}

Figure \ref{leptons} shows (a) the transverse momentum and (b) rapidity of
the highest transverse momentum charged lepton for signal and background processes.
The \( W+ \)jets background is very similar to the signal in these distributions.
Leptons from the \( t\overline{t} \) background are slightly softer and more
central. Figure \ref{leptons}(c) shows the missing transverse momentum. Again,
the \( t\overline{t} \) background is slightly softer than the other two samples. 

All leptons in an event are then combined one-by-one to give, if possible, a
reconstructed \( W \) boson (to within a twofold ambiguity due to the unknown
\( z \) component of the neutrino momentum). The transverse momentum of all
these \( W \) candidates is shown in Figure \ref{leptons}(d). The signal has
a harder distribution than both backgrounds. A selection cut is applied at 320
GeV on this distribution and in the case that more than one candidate is present,
that with the highest transverse momentum is used.

\subsection{The Hadronic \protect\( W\protect \) Decay}

\begin{figure}
{\par\centering \resizebox*{1\textwidth}{!}{\includegraphics{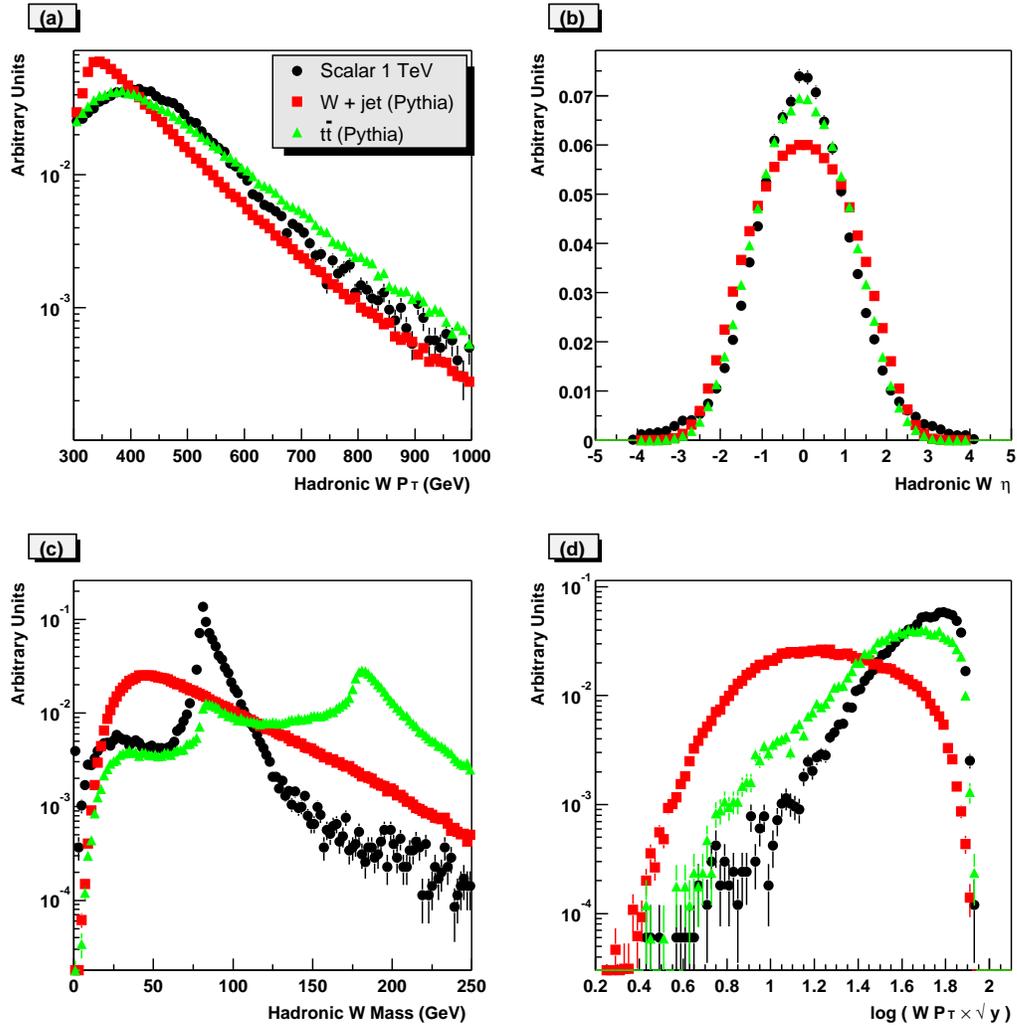}} \par}

\caption{\label{hadw}Kinematic variables for the hadronically decaying \protect\( W\protect \)
candidate. (a) \protect\( p_{T}\protect \), (b) \protect\( \eta \protect \),
(c) Invariant mass (d) \protect\( p_{T}\sqrt{y}\protect \) . The area under
the histograms is set to unity to allow comparison of the shapes.}
\end{figure}

Figure \ref{hadw}(a) shows the transverse momentum and (b) the pseudorapidity
(\( \eta  \)) of the highest transverse momentum jet in the remaining signal
and background samples. Jet finding is performed with the inclusive \( k_{T} \)
algorithm \cite{kt}, and the \( E \) recombination scheme is used throughout.
To reconstruct the \( W \) mass, the highest transverse momentum jet within
the region \( |\eta |<4 \) is selected. In the \( E \) recombination scheme
the candidate \( W \) mass, \( M_{J} \) is then the invariant mass of this
jet. Figure \ref{hadw}(c) shows this distribution, with \( W \) mass peaks
visible in the signal and in the \( t\overline{t} \) sample, and a top mass
peak also visible in the \( t\overline{t} \) events. Cuts are applied at \( p_{T}>320 \)
GeV, and \( 70 \)~GeV\( <M_{J}<90 \)~GeV. The results (after this cut and
the leptonic cuts) are shown in the second and third rows of Table \ref{cuts}. 

The jet is next forced to decompose into two subjets. The possibility of using
subjets to reconstruct massive particles decaying to hadrons has been discussed
previously \cite{mike}. In this analysis we develop a new technique. The extra
pieces of information gained from the subjet decomposition are the \( y \)
cut at which the subjets are defined and the four-vectors of the subjets. For
a genuine \( W \) decay the expectation is that the scale at which the jet
is resolved into subje\textit{\emph{ts (i.e.}} \( yp_{T}^{2} \)) will be \( {\cal {O}}(M_{W}^{2}) \).
The distribution of \( \log (p_{T}\sqrt{y}) \) is shown in Figure \ref{hadw}(d).
The scale of the splitting is indeed high in the signal and softer in the \( W+ \)
jets background, where the hadronic \( W \) is in general a QCD jet rather
than a genuine second \( W \). A cut is applied at \( 1.6<\log (p_{T}\sqrt{y})<2.0 \).
The effect of this cut is shown in the fourth row of the table. Whilst this
is a powerful cut for reducing the \( W+ \) jets background, the effect on
the \( t\overline{t} \) background, which more often contains two real \( W \)
bosons, is less marked.

\subsection{The Hadronic Environment\label{HadEnv}}

To further reduce backgrounds, cuts must be applied to characteristics of the
event other than those directly related to the decaying \( W \) bosons.

\paragraph*{Top quark veto}

\begin{figure}
{\par\centering \resizebox*{1\textwidth}{!}{\includegraphics{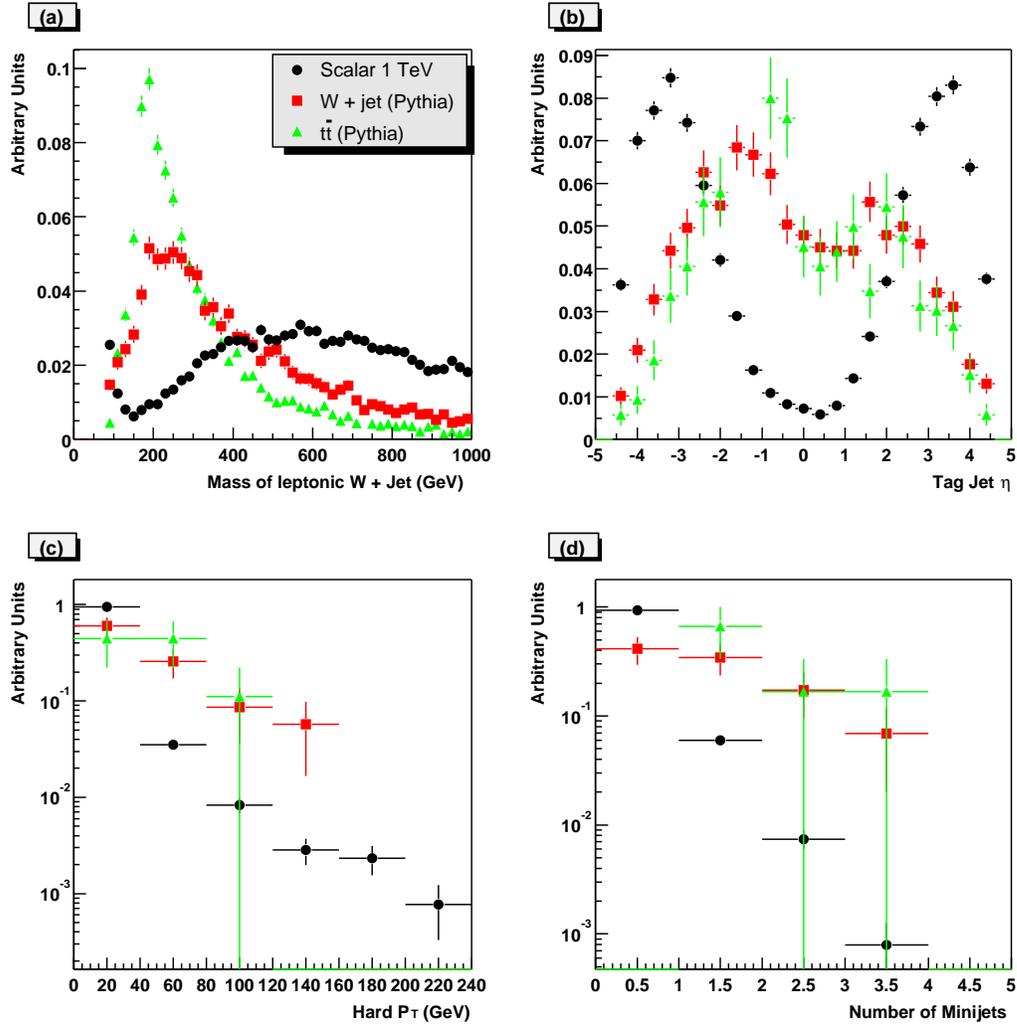}} \par}

\caption{\label{hfs}(a) The mass distribution for the leptonic W candidate combined
seperately with all other jets in the event other than the hadronic \protect\( W\protect \)
candidate. (b) The rapidity distribution for tag jets (see text). (c) The transverse
momentum distribution for the \protect\( WW+\protect \) tag jets system. (d)
The number of minijets (see text).}
\end{figure}

In the remaining \( t\overline{t} \) events containing a genuine leptonic \( W \),
the \( W \) will combine with a jet other than the hadronic \( W \) candidate
to give a mass close to the top mass. This mass distribution for the leptonic
\( W \) candidate combined separately with each such jet in the event is shown
in Figure \ref{hfs}(a). The top peak is clearly visible in the \( t\overline{t} \)
sample. Any event with a mass in the region 130 GeV \( <M_{wj}< \) 240 GeV
is rejected. A similar distribution (not shown) is obtained by combining the
hadronic \( W \) candidate with other jets in the event, and the same cut is
applied. In combination these cuts are decribed as a {}``top quark veto{}'',
and their effect is shown row five of Table \ref{cuts}.

\paragraph*{Tag jets}

In the \( WW \) scattering process the bosons are radiated from quarks in the
initial state (see Figure \ref{Feynman}(a)). The quark from which the boson
is radiated will give a jet at high rapidity (i.e. close to the direction of
the hadron from which it emerged). These jets are not in general present in
the background processes and demanding their presence is therefore a powerful
tag of the signal \cite{tagjets}. In this analysis we define a {}``tag jet{}''
as follows. The event is divided into three regions of rapidity: {}``forward{}'',
i.e. forward of the most forward \( W \); {}``backward{}'', i.e. backward
of the most backward \( W \); and {}``central{}''\textit{,} i.e. the remaining
region, which includes both \( W \) candidates. A forward (backward) tag jet
is defined as the highest transverse energy jet in the forward (backward) region.
In Figure \ref{hfs}(b) the rapidity distribution of the tag jets with \( p_{T}>20 \)
GeV is shown. Signal events display an enhancement at high \( |\eta | \) and
a suppression at low \( |\eta | \), in dramatic contrast to the background
processes, where most jets are central. For an event to be retained it must
have a tag jet in both the forward and backward regions satisfying \( p_{T}>20 \)
GeV, \( E>300 \) GeV and \( 4.5>|\eta |>2 \). The result of imposing this
cut is shown in row six of Table \ref{cuts}. The background is reduced by a
factor of around fifty, at the cost of a loss less than two thirds of the signal.

\paragraph*{Hard \protect\( p_{T}\protect \)}

Figure \ref{hfs}(c) shows the \( p_{T} \) distribution for the {}``hard scattering{}''
system comprising the two tags jets and the two \( W \) candidates. For events
surviving the cuts so far, the background events have a harder spectrum than
the signal, since in the signal events this system is the complete result of
a scattering between colinear partons, whereas in the backgrounds extra jets
from hard QCD radiation may be picked up and/or missed. An upper cut is applied
at 50 GeV, and the results are shown in row seven of Table \ref{cuts}.

\paragraph{Minijet Veto}

Finally, a cut which has been employed before in similar analyses \cite{TDR,zepp,barger}
exploits the fact that for signal events no colour is exchanged between the
quarks which radiate the \( W \) bosons and the jets which are produced by
the hadronically decaying \( W \). This leads to a suppression of QCD radiation
in the central region in the signal with respect to the background. However,
significant activity is expected in all classes of event due to remnant-remnant
interactions ({}``underlying event{}''). This activity can produce additional
(mini)jets, and so it is important to choose a cut on additional jet activity
which is robust against the large uncertainties in current understanding of
the underlying event at the LHC. In this analysis minijets are defined as all
jets apart from the hadronic \( W \) candidate with \( |\eta |<2 \). Events
are vetoed if the number of minijets with \( p_{T}>15 \) GeV is greater than
one. The distribution of the number of jets satisfying these demands is shown
in Figure \ref{hfs}(d). The result of applying this cut is shown in row eight
of Table \ref{cuts}. This cut is discussed further in Section \ref{UE}.

\section{Analysing the signal }

\subsection{Efficiency and Event Numbers}

\begin{figure}
{\par\centering \resizebox*{1\textwidth}{0.9\textheight}{\includegraphics{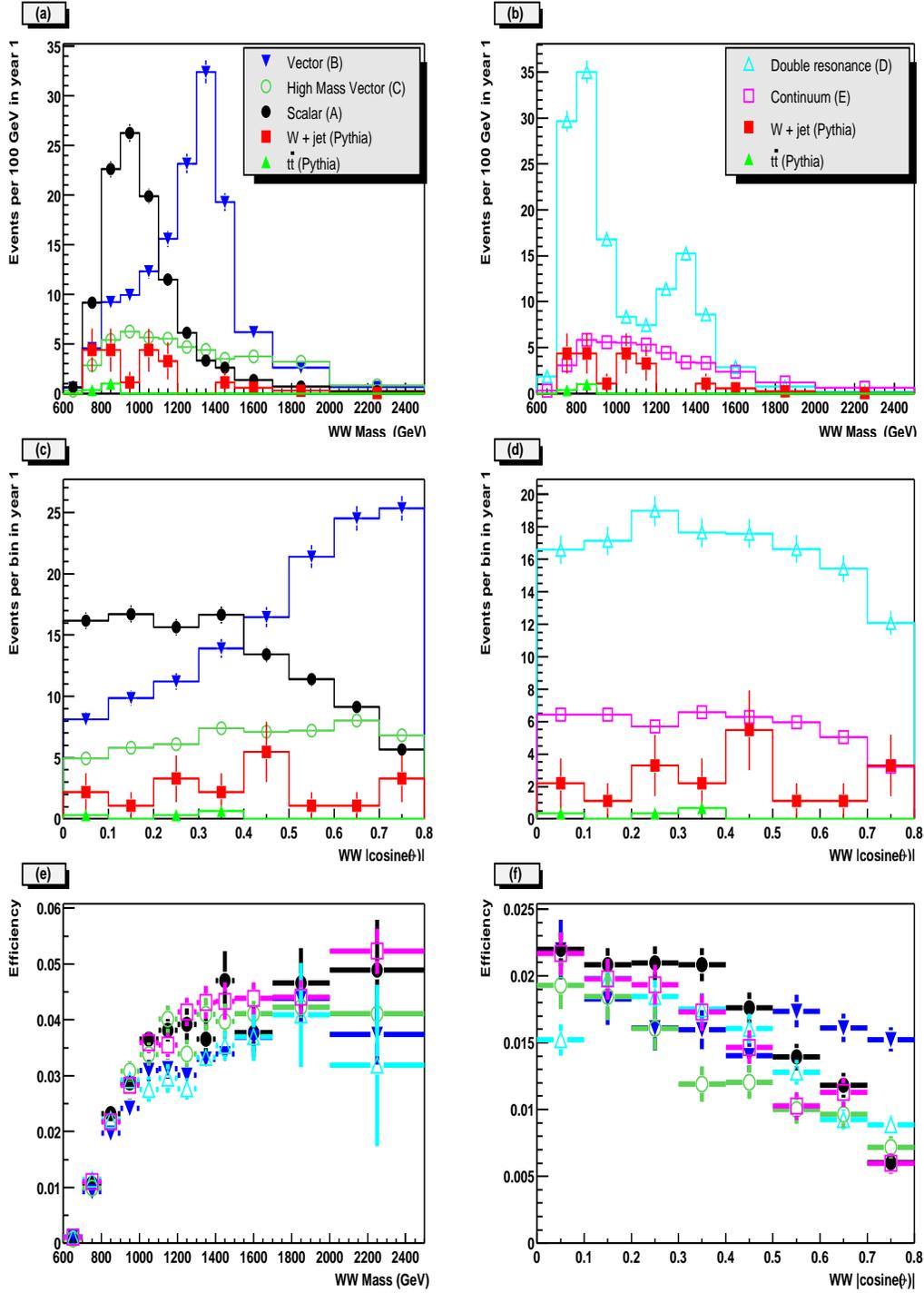}} \par}

\caption{\label{effs}(a,b) Distribution of the reconstructed \protect\( WW\protect \)
mass for signals and backgrounds separately. (c,d) Distribution of \protect\( |\cos \theta ^{*}|\protect \),
the absolute value of the cosine of the centre-of-mass scattering angle for
signals and backgrounds seperately. (e) Efficiency for signal events as a function
of the true \protect\( M_{WW}\protect \) and (f) \protect\( |\cos \theta ^{*}|\protect \).
The errors reflect the statistics which would be obtained after approximately
one year of running at the LHC, i.e. 100 fb\protect\( ^{-1}\protect \).}
\end{figure}

Having applied the cuts described in the previous section, the \( WW \) mass
distribution obtained is shown in Figure \ref{effs}(a) and (b) for all five
signal samples discussed above. The resolution obtained in this variable is
around 10 GeV, before any detector smearing. The efficiency is shown as a function
of the true \( WW \) mass in (e). It rises from zero to 6\% between 500 GeV
and 1.5 TeV, and is flat above this value. This efficiency includes the branching
ratio for semileptonic \( W \) decays of around 15\% . Excluding the branching
ratio, the efficiency is around 40\%.

A key variable for distinguishing between scalar and vector resonances is the
angular distribution of the scattered \( W \) pair in the \( WW \) centre-of-mass
system. In Figure \ref{effs}(c) and (d) the distribution of \( |\cos \theta ^{*}| \)
is shown, where \( \theta ^{*} \) is the angle between the scattered \( W \)
and the incoming \( W \) direction, in the \( WW \) centre-of-mass frame.
In (f) the efficiency is shown. The efficiency is very dependent on the mass
distribution, since for the same transverse momentum, high scattering angles
have high mass. This means that the transverse momentum cuts bias this distribution.
However, this bias is well understood and could be corrected for in a final
measurement using a two-dimensional correction in mass and angle regardless
of the input distribution.
\vspace{0.3cm}

\subsection{Simulated Measurement}

\begin{figure}
{\par\centering \resizebox*{1\textwidth}{0.9\textheight}{\includegraphics{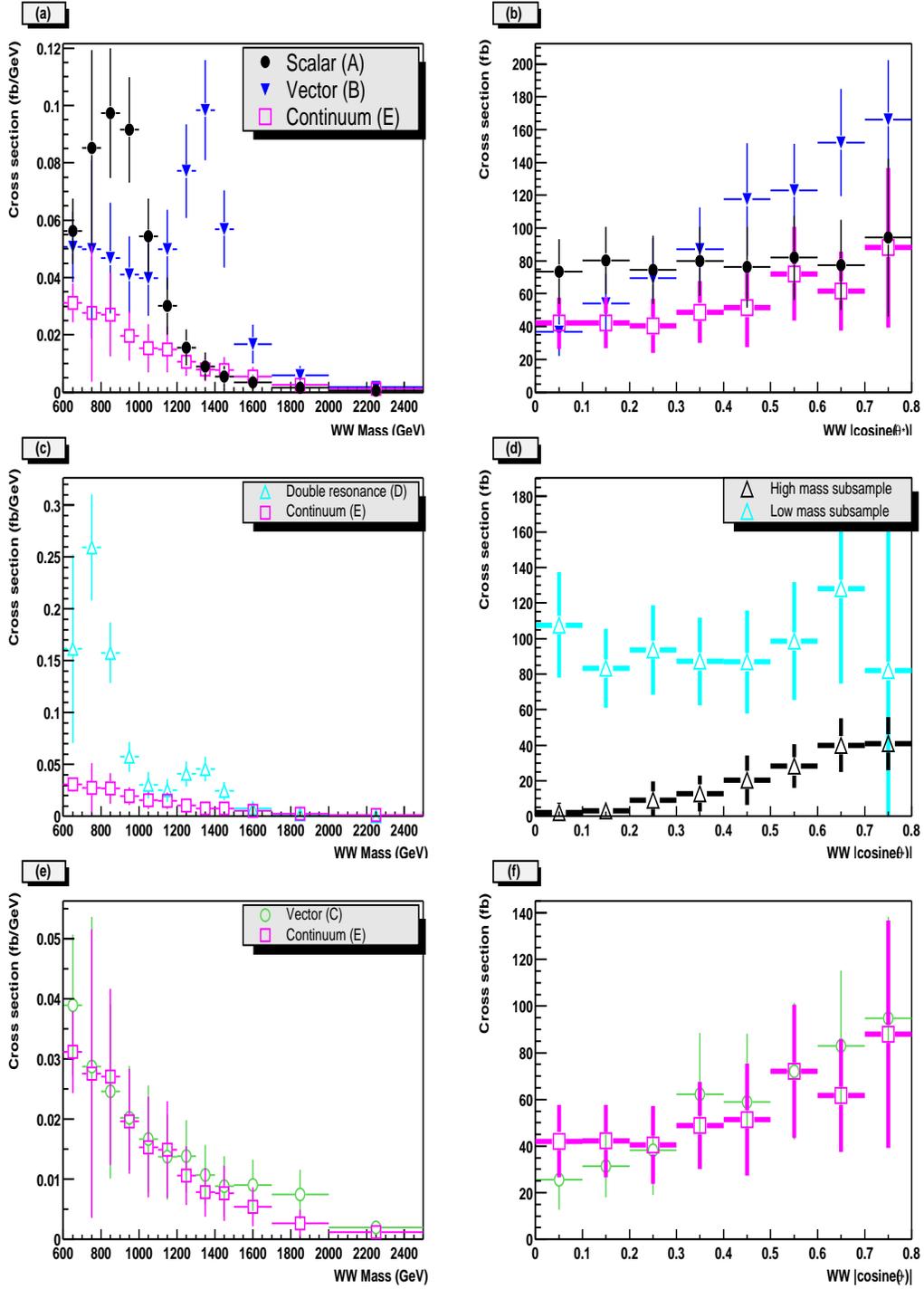}} \par}

\caption{\label{final}Measurement expectation after 100 fb\protect\( ^{-1}\protect \)
of LHC luminosity at 14 TeV cm energy. (a,c,e) \protect\( d\sigma /dM_{WW}\protect \)
and (b,d,f) \protect\( d\sigma /d|\cos \theta ^{*}|\protect \). (d) shows \protect\( d\sigma /d|\cos \theta ^{*}|\protect \)
for the high and low mass subsamples for the double resonance model, separated
by a cut at 1200 GeV. }
\end{figure}

If it is assumed that the backgrounds can be well constrained from developments
in theory, measurements at the Tevatron and HERA over the next few years, and
measurements at the LHC in other kinematic regions, then the statistical error
on an extraction of the \( M_{WW} \) and \( |\cos \theta ^{*}| \) distributions
can be estimated by adding the statistical errors on the signal and background
distributions in quadrature. Under this assumption, a simulation of an expected
measurement of the differential cross-section \( d\sigma /dM_{WW} \) after
100~fb\( ^{-1} \) of LHC luminosity is shown in Figure \ref{final}(a), (c)
and (e). The scenarios containing resonances are distinguishable above the background,
and are also distinguishable from each other due to their different resonant
masses. In Figure \ref{final}(c) the double resonance sample (D) is shown,
with two peaks clearly measured. Also shown (in all three figures) is the continuum
model (E).

The expected measurement of the differential cross-section \( d\sigma /d|\cos \vartheta ^{*}| \)
after 100 fb\( ^{-1} \) of LHC luminosity is shown in Figure \ref{final}(b),(d)
and (f) for \( M_{WW}>750 \) GeV. The intermediate mass vector and scalar resonances
have the expected behaviour, with the vector rising towards high \( |\cos \theta ^{*}| \)
and the scalar being flat. In Figure \ref{final}(d) the distribution for the
double resonance model is shown in two mass bins: \( 750<M_{WW}<1200 \) GeV
and \( M_{WW}>1200 \) GeV. With the high statistics generated (corresponding
to a very high integrated luminosity), the lower mass resonance can be seen
to be a scalar whilst the higher mass is a vector. However, within the simulated
errors the measurement of the spin of the lower mass resonance would be marginal.
\vspace{0.3cm}

\begin{table}
{\centering \begin{tabular}{|c|c|c|c|c|c|}
\hline 
 {\small Cuts}&
{\small Efficiency}&
{\small Signal}&
{\small \( t\overline{t} \) }&
{\small \( W+ \)Jets}&
{\small Sig/B}\\
{\small }&
{\small }&
{\small \( \sigma  \) (fb)}&
{\small \( \sigma  \) (fb)}&
{\small \( \sigma  \) (fb)}&
{\small }\\
\hline 
\textbf{\small Generated} {\small }&
{\small A:100\%}&
{\small 72}&
\multicolumn{2}{|c|}{{\small Pythia }}&
{\small \( 8.7\times 10^{-4} \)}\\
{\small }&
{\small B:100\%}&
{\small 104}&
{\small 18,000}&
{\small 65,000}&
{\small \( 1.3\times 10^{-3} \)}\\
{\small }&
{\small C:100\%}&
{\small 44}&
\multicolumn{2}{|c|}{{\small Herwig }}&
{\small \( 5.3\times 10^{-4} \)}\\
{\small }&
{\small D:100\%}&
{\small 113}&
{\small 14,000}&
{\small 53,000}&
{\small \( 1.4\times 10^{-3} \)}\\
{\small }&
{\small E:100\%}&
{\small 47}&
{\small }&
{\small }&
{\small \( 5.0\times 10^{-4} \)}\\
\hline 
{\small \( p_{T} \) (Lep. \( W \) )\( >320 \) GeV}&
{\small A:11\%}&
{\small 8.2}&
\multicolumn{2}{|c|}{{\small Pythia }}&
{\small \( 1.5\times 10^{-3} \)}\\
{\small and}&
{\small B:11\%}&
{\small 11}&
{\small 910}&
{\small 4400}&
{\small \( 2.1\times 10^{-3} \)}\\
{\small \( p_{T} \) (Had. \( W \)) \( >320 \) GeV}&
{\small C:10\%}&
{\small 4.4}&
\multicolumn{2}{|c|}{{\small Herwig }}&
{\small \( 8.3\times 10^{-4} \)}\\
{\small }&
{\small D:10\%}&
{\small 11}&
{\small 750}&
{\small 3600}&
{\small \( 2.1\times 10^{-3} \)}\\
{\small }&
{\small E:10\%}&
{\small 4.7}&
{\small }&
{\small }&
{\small \( 8.8\times 10^{-4} \)}\\
\hline 
 {\small 70 GeV \( <M \)(Had. \( W \) )}&
{\small A:6.7\%}&
{\small 4.8}&
\multicolumn{2}{|c|}{{\small Pythia }}&
{\small \( 6.3\times 10^{-3} \)}\\
{\small \( <90 \) GeV}&
{\small B:6.2\%}&
{\small 6.4}&
{\small 56}&
{\small 700}&
{\small \( 8.4\times 10^{-3} \)}\\
{\small }&
{\small C:5.8\%}&
{\small 2.6}&
\multicolumn{2}{|c|}{ {\small Herwig}}&
{\small \( 3.4\times 10^{-3} \)}\\
{\small }&
{\small D:5.6\%}&
{\small 6.3}&
{\small 52}&
{\small 480}&
{\small \( 8.3\times 10^{-3} \)}\\
{\small }&
{\small E:5.8\%}&
{\small 2.7}&
{\small }&
{\small }&
{\small \( 3.6\times 10^{-3} \)}\\
\hline 
{\small \( 1.6<\log (p_{T}\times \sqrt{y} \) \( )<2.0 \)}&
{\small A:4.7\%}&
{\small 3.4}&
\multicolumn{2}{|c|}{ {\small Pythia}}&
{\small \( 3.2\times 10^{-2} \)}\\
{\small }&
{\small B:4.4\%}&
{\small 4.5}&
{\small 28}&
{\small 78}&
{\small \( 4.3\times 10^{-2} \)}\\
{\small }&
{\small C:4.1\%}&
{\small 1.8}&
\multicolumn{2}{|c|}{{\small Herwig }}&
{\small \( 1.7\times 10^{-2} \)}\\
{\small }&
{\small D:4.0\%}&
{\small 4.5}&
{\small 27}&
{\small 66}&
{\small \( 4.3\times 10^{-2} \)}\\
{\small }&
{\small E:4.1\%}&
{\small 1.9}&
{\small }&
{\small }&
{\small \( 1.8\times 10^{-2} \)}\\
\hline 
{\small Top quark veto}&
{\small A:4.3\%}&
{\small 3.1}&
\multicolumn{2}{|c|}{{\small Pythia }}&
{\small \( 5.6\times 10^{-2} \)}\\
{\small (see text)}&
{\small B:4.0\%}&
{\small 4.2}&
{\small 3.2}&
{\small 52}&
{\small \( 7.5\times 10^{-2} \)}\\
{\small }&
{\small C:3.8\%}&
{\small 1.7}&
\multicolumn{2}{|c|}{{\small Herwig }}&
{\small \( 3.0\times 10^{-2} \)}\\
{\small }&
{\small D:3.6\%}&
{\small 4.1}&
{\small 3.4}&
{\small 43}&
{\small \( 7.3\times 10^{-2} \)}\\
{\small }&
{\small E:3.8\%}&
{\small 1.8}&
{\small }&
{\small }&
{\small \( 3.2\times 10^{-2} \)}\\
\hline 
{\small Tag jets }&
{\small A:1.6\%}&
{\small 1.1}&
\multicolumn{2}{|c|}{{\small Pythia }}&
{\small 2.7}\\
{\small \( p_{T}>20 \) GeV, \( E>300 \) GeV}&
{\small B:1.5\%}&
{\small 1.6}&
{\small 0.030}&
{\small 0.38}&
{\small 3.8}\\
{\small (see text)}&
{\small C:1.4\%}&
{\small 0.63}&
\multicolumn{2}{|c|}{{\small Herwig }}&
{\small 1.5}\\
{\small }&
{\small D:1.3\%}&
{\small 1.5}&
{\small 0.082}&
{\small 0.42}&
{\small 3.6}\\
{\small }&
{\small E:1.4\%}&
{\small 0.67}&
{\small }&
{\small }&
{\small 1.6}\\
\hline 
{\small Hard \( p_{T}<50 \) GeV}&
{\small A:1.5\%}&
{\small 1.1}&
\multicolumn{2}{|c|}{ {\small Pythia}}&
{\small 3.2}\\
{\small }&
{\small B:1.5\%}&
{\small 1.5}&
{\small 0.020}&
{\small 0.32}&
{\small 4.5}\\
{\small }&
{\small C:1.4\%}&
{\small 0.61}&
\multicolumn{2}{|c|}{ {\small Herwig}}&
{\small 1.8}\\
{\small }&
{\small D:1.3\%}&
{\small 1.4}&
{\small 0.048}&
{\small 0.37}&
{\small 4.3}\\
{\small }&
{\small E:1.4\%}&
{\small 0.65}&
{\small }&
{\small }&
{\small 1.9}\\
\hline 
{\small Minijet veto }&
{\small A:1.5\%}&
{\small 1.1}&
\multicolumn{2}{|c|}{{\small Pythia }}&
{\small 4.3}\\
{\small \( p_{T}>15 \) GeV, see text}&
{\small B:1.5\%}&
{\small 1.5}&
{\small 0.013}&
{\small 0.24}&
{\small 6.0}\\
{\small }&
{\small C:1.4\%}&
{\small 0.61}&
\multicolumn{2}{|c|}{{\small Herwig }}&
{\small 2.4}\\
{\small }&
{\small D:1.3\%}&
{\small 1.4}&
{\small 0.048}&
{\small 0.36}&
{\small 5.6}\\
{\small }&
{\small E:1.4\%}&
{\small 0.65}&
{\small }&
{\small }&
{\small 2.6}\\
\hline 
\end{tabular}\small \par}

\caption{\label{cuts}The effect of cuts on the signal and background samples. A: 1
TeV scalar, B: 1.4 TeV Vector, C: 2 TeV Vector, D: Double Resonance and E: Continuum.}
\end{table}

\section{The Underlying Event\label{UE}}

One of the more uncertain aspects of the analysis is the understanding of the
so-called {}``underlying event{}''. This is defined here as particle and energy
flow in the event associated with the same proton-proton interaction but incoherent
with the \( W \) production process. Hence we explicitly exclude from our definition
the effects of multiple \( pp \) interactions in the same bunch crossing, any
detector effects such as those associated with noise or pile-up, and hard QCD
radiation associated directly with the hard scatter. The first two of these
are not simulated here and controlling and understanding them requires detailed
experimental work. The third is simulated to leading-logarithmic accuracy in
both \textsc{Pythia} and \textsc{Herwig}. While this simulation should and probably
will be improved in the future, for now it is considered adequate.

The remaining activity can be characterised as interactions between the proton
remnant systems. It is important because it is largely independent of the hard
scattering process, and therefore contributes to minijet activity in both signal
and background, degrading the effectiveness of the minijet veto. In addition,
underlying event activity contributes to the observed \( W \) width and the
position of the mass peaks in a highly model-dependent way.

\begin{figure}
{\par\centering \resizebox*{1\textwidth}{!}{\includegraphics{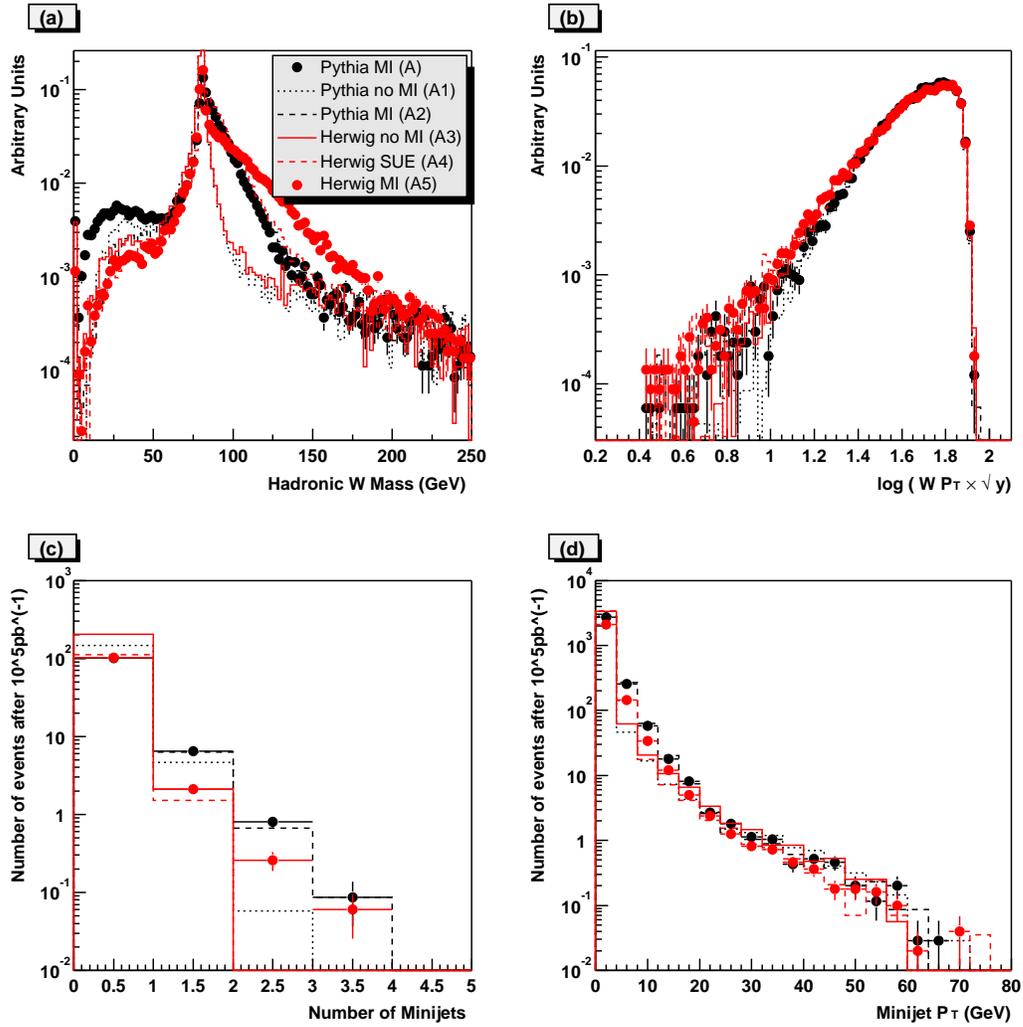}} \par}

\caption{\label{mi}The effect of the underlying event. (a) Hadronic \protect\( W\protect \)
mass, (b) \protect\( p_{T}\sqrt{y}\protect \) , (c) Number of minijets and
(d) the \protect\( p_{T}\protect \) distribution of the minijets.}
\end{figure}

In Figure \ref{mi}(a) and (b) the jet mass distribution and the \( \log (p_{T}\sqrt{y}) \)
are shown again (as in Figure \ref{hadw}(c) and (d)) for the signal events
(1 TeV resonance, sample A) using our default underlying event model. In addition,
several other underlying event models are shown. In \textsc{Pythia,} we turn
off multiparton interactions (sample A1), and turn on the default model (sample
A2) which has a \( p_{T}^{min} \) of 2.89 GeV at LHC energies. Also shown are
three samples of 1 TeV Higgs events generated using \noun{herwig}. These have
no underlying event (sample A3), soft underlying event (sample A4) and multiparton
interactions generated with fixed \( p_{T}^{min}=3.0 \) GeV (sample A5)\cite{jimmy}.
The width of the \( W \) mass peak is much greater in general for those samples
which include an underlying event. Whilst the \noun{pythia} multiparton interaction
models and the \textsc{Herwig} soft underlying event are fairly consistent with
each other, the \textsc{Herwig} multiparton interaction model gives a very different
distribution. However, the \( \log (p_{T}\sqrt{y}) \) is similar for all models,
implying that this cut should be robust against such uncertainties.

For the same samples, the minijet \( p_{T} \) distribution and the number of
minjets passing the 15 GeV cut, which we introduced in the analysis of Section
\ref{HadEnv}, are shown in Figure \ref{mi}(c) and (d), with absolute normalisation.
In contrast to the \( W \) mass distribution, in these distributions the \textsc{Herwig}
multiparton interaction model is close to the \noun{pythia} multiparton models,
whereas the soft underlying event model is closer to the models without underlying
event. The \( p_{T} \) distribution is very steeply falling, and is sensitive
to the underlying event below around 20 GeV. Thus, there is sensitivity in the
number of jets at 15 GeV, and this would become worse for lower choices of cut.
Lowering the cut further without introducing large uncertainties requires a
better knowledge of the underlying event than is currently available.

If the no underlying event model is used in \noun{pythia (}sample A1), the
signal/background for the scenario A is 8.0. However, for all other cases (models
A, A2-A5) the ratio is between 2.5 and 4.0. Data from the Tevatron and photoproduction
at HERA (see for example \cite{midata} and references therein), strongly disfavour
models without underlying event (A1, A3) and are generally more consistent with
the other models considered here (though none provides a perfect description).
However, further work is needed on constraining these models to improve confidence
in the extrapolation to the LHC. At present a systematic error of 40-50\% would
have to be assigned to the measurement from this source alone.

\section{Summary and Conclusions}

A major goal of the LHC is to extract the \( WW\rightarrow WW \) cross-section
as accurately as possible to the highest centre-of-mass energies in order to
shed light on the nature of electroweak symmetry breaking. We have performed
a study of the \( WW\to WW \) scattering cross-section in the scenario that
there is no new physics below the TeV scale using the formalism of the Electroweak
Chiral Lagrangian extended by the imposition of unitarity constraints. Two different
unitarisation protocols are used: Pad\'e and N/D. These protocols determine
the behaviour of the scattering cross-section into the TeV regime and they typically
predict the emergence of new vector and/or scalar resonances. We have performed
a detailed comparison of these two unitarisation methods. 

We have implemented the physics of the unitarised Electroweak Chiral Lagrangian
in a realistic general-purpose Monte Carlo (\textsc{Pythia}). The semi-leptonic
decay mode of the final state \( W \) pair has been studied at the final state
particle level with detector acceptance cuts but no smearing. We have considered
five different physics scenarios which are representative of the different types
of physics which we might reasonably expect at the LHC. The principal backgrounds
come from \( W+ \) jet and \( t\overline{t} \) production, and we consider
these backgrounds using both the \noun{pythia} and \noun{herwig} Monte Carlos.
A new method for identifying hadronically decaying \( W \) bosons is introduced
which we expect to be useful more generally in the identification of hadronically
decaying massive particles which have energy large compared to their mass. Other
new features include a top quark veto and a cut on the transverse momentum of
the hard subsystem. In addition, the established tag jet and minijet veto cuts
are applied. The results are cross-checked with \textsc{Herwig} using a simulation
of a 1 TeV Higgs boson for the signal. The effect of uncertainties in the underlying
event leads to a model dependent systematic error of 40-50\%. New data from
Tevatron and HERA should help to reduce this before the LHC turns on. 

The results compare very well with previous Higgs search studies in the semi-leptonic
channel. Over a wide range of parameter space signal/background ratios of greater
than unity can be obtained, and the cross-section can be measured differentially
in the \( WW \) centre-of-mass energy within one year of high luminosity LHC
running (100 fb\( ^{-1} \)). Vector and scalar resonances up to around 1.5
TeV may well be observable, and their spins measureable. Detailing the exact
regions of sensitivity, as well as verifying the improvements in signal/background
arising from the new cuts, requires a more detailed simulation of the LHC general
purpose detectors.

\newpage
\subsection*{Acknowledgements}

Thanks to Anahita New, Jos\'e Oller and Graham Shaw for their valuable contributions.
This work was supported by the UK Particle Physics and Astronomy Research Council
(PPARC).

\end{document}

%% file: math.tex
\def \f{\frac}

\newcounter{indfig}
\def\lapproxeq{\lower .7ex\hbox{$\;\stackrel{\textstyle<}{\sim}\;$}}
\def\gapproxeq{\lower .7ex\hbox{$\;\stackrel{\textstyle>}{\sim}\;$}}

\def\re{{\rm Re}}
\def\im{{\rm Im}}